\documentclass[11pt]{article}
\usepackage{verbatim,amsmath,amssymb}
\usepackage{epsfig,float,color}
\usepackage{epstopdf}
\usepackage{geometry}
\usepackage{setspace}
\usepackage{wrapfig}
\usepackage{hyperref}
\usepackage[utf8]{inputenc}
\usepackage{graphicx}
\usepackage{flushend}
\usepackage{tikz,pgfplots}
\usepackage[linesnumbered,ruled]{algorithm2e}

\usepackage[font={footnotesize}]{caption}
\usepackage[font={footnotesize}]{subcaption}
\usepackage{footnote}
\usepackage{multicol}
\usepackage{multirow}
\usepackage{enumitem}   
\usepackage{soul}
\usepackage{framed}
\usepackage[titletoc,title]{appendix}
\usepackage{bm}
\usepackage{siunitx}
\usepackage{cite}
\usepackage[makeroom]{cancel} 
\usepackage{natbib}
\usepackage{hyperref}
\usepackage{bigints} 

\definecolor{darkblue}{rgb}{0,0,1}
\hypersetup{pdftex=true, colorlinks=true, breaklinks=true, linkcolor=blue, menucolor=blue, citecolor=blue, urlcolor=blue}

\newcommand{\pd}[2]{\frac{\partial #1}{\partial #2}}

\usepackage{tikz}
\usetikzlibrary{positioning, arrows.meta}
\tikzset{%
	myarrow/.style = {-Stealth, shorten >=5pt}
}

\newenvironment{rcases}
{\left.\begin{aligned}}
	{\end{aligned}\right\rbrace}

\newenvironment{lcases}
{\left\lbrace\begin{aligned}}
	{\end{aligned}\right.}

\usepackage{listings}
\definecolor{LightCyan}{rgb}{0.88,1,1}
\definecolor{mygreen}{RGB}{28,172,0} 
\definecolor{mylilas}{RGB}{170,55,241}
\begin{document}
	
	\begin{center}
		\Large{\bf{\texttt{\textbf{HoneyTop90}}: A 90-line MATLAB code for topology optimization using honeycomb tessellation}}\\
		
	\end{center}
	
	\begin{center}
		
		\large{P. Kumar $^{\star,\,\dagger,}$\footnote{pkumar@mae.iith.ac.in; prabhatkumar.rns@gmail.com}}
		\vspace{4mm}
		
		\small{{$\star$}\textit{Department of Mechanical and Aerospace Engineering, Indian Institute of Technology Hyderabad, 502285, India}}
		\vspace{4mm}
		
		\small{\textit{$\dagger$ \textit{Department of Mechanical Engineering, Indian Institute of Science,
					Bengaluru, 560012, Karnataka, India}}}\\
		\vspace{4mm}
		
		Published\footnote{This pdf is the personal version of an article whose final publication is available at \href{https://link.springer.com/article/10.1007/s11081-022-09715-6}{Optimization and Engineering}}\,\,\,in \textit{Optimization and Engineering}, 
		\href{https://link.springer.com/article/10.1007/s11081-022-09715-6}{DOI:10.1007/s11081-022-09715-6} \\
		Submitted on 01~September 2021, Revised on 15~February 2022, Accepted on 17~March 2022
		
	\end{center}
	
	\vspace{1mm}
	\rule{\linewidth}{.15mm}
	{\bf Abstract:}
	This paper provides a simple, compact and efficient 90-line pedagogical MATLAB  code  for topology optimization using hexagonal elements (honeycomb tessellation). Hexagonal elements provide nonsingular connectivity between two juxtaposed elements and, thus, subdue checkerboard patterns and point connections inherently from the optimized designs. A novel approach to generate honeycomb tessellation is proposed. The element connectivity matrix and corresponding nodal coordinates array are determined in 5 (7) and 4 (6) lines, respectively.  Two additional lines for the meshgrid generation are required for an even number of elements in the vertical direction. The code takes a fraction of a second to generate meshgrid information for the millions of hexagonal elements. Wachspress shape functions are employed for the finite element analysis, and compliance minimization is performed using the optimality criteria method. The provided Matlab code and its extensions are explained in detail. Options to run the optimization with and without filtering techniques are provided. Steps to include different boundary conditions, multiple load cases, active and passive regions, and a Heaviside projection filter are also discussed. The code is provided in Appendix~\ref{app:MATLABcode}, and it can also be downloaded along with supplementary materials from \url{https://github.com/PrabhatIn/HoneyTop90}.\\
	

	\vspace{-4mm}
	\rule{\linewidth}{.15mm}
	
	\section{Introduction}
Topology optimization (TO), a design technique, determines an optimized material distribution within a specified design domain with known boundary conditions by extremizing an objective for the given physical and geometrical constraints \citep{sigmund2013topology}. In a typical continuum optimization setting, the design domain is parameterized by either quadrilateral or polygonal finite elements (FEs), and the associated boundary value problems are solved. Each FE is assigned a design variable $\rho\in[0,\,1]$. $\rho =1$ and $\rho=0$ indicate the solid and the void states of the element respectively.

There exist numerous  pedagogical TO MATLAB codes online in \cite{sigmund200199,suresh2010199,challis2010discrete,huang2010evolutionary,andreassen2011efficient,saxena2011topology,talischi2012polytop,wei201888,ferrari2020new,picelli2020101} that can help a user to learn and explore various optimization techniques. In addition, one may refer to the articles by \cite{han2021efficient,wang2021comprehensive} for a comprehensive discussion on various TO educational codes.  A simple, compact and efficient educational code using pure (regular) hexagonal finite elements (honeycomb tessellation)  with Wachspress shape functions however cannot be found in the current state-of-the-art of TO. Such  codes can find importance for  newcomer students to learn, explore, realize and visualize  the characteristics of  hexagonal FEs in a TO framework with a minimum effort. Honeycomb tessellation offers nonsingular geometric connectivity and thus, circumvents checkerboard patterns and single-point connections inherently from the optimized designs \citep{saxena2003honeycomb,saxena2007honeycomb,langelaar2007use,talischi2009honeycomb,saxena2011topology,kumar2017_diss}. In addition, as per \cite{sukumar2004conforming}, polygonal/hexagonal FEs provide better accuracy in numerical solutions and are suitable for modeling of polycrystalline materials.

An approach with hexagonal FEs  is presented by \cite{saxena2011topology}, and the author also shares the related MATLAB code. However, a new reader may not find the code generic primarily because it does not detail how to generate the honeycomb tessellation for a given problem.  \cite{talischi2012polymesher} provide a MATLAB code
 to generate polygonal mesh using implicit description of the domain and the centroidal Voronoi diagrams. The code is suitable to parameterize arbitrary geometrical shapes using polygonal elements, however it requires many subroutines and involved processes and thus, a newcomer may find difficulties to learn and explore with that code. \cite{talischi2012polytop} use the polygonal meshing method \citep{talischi2012polymesher} in their TO approach. TO approaches using the polymesher code \citep{talischi2012polymesher} can also be found in \cite{sanders2018polymat,giraldo2021polystress}. The motif herein is to provide a simple, compact, efficient and hands-on pedagogical MATLAB code with hexagonal elements such that a user can readily: (A) generate hexagonal FEs, (B) obtain the corresponding element connectivity matrix and nodal coordinates array and (C) perform FE analysis and TO and also, visualize the intermediate design evolution of the optimization in line with \citep{andreassen2011efficient,ferrari2020new}. (A) is expected to significantly reduce the learning time for newcomers in TO using honeycomb tessellation, while this in association with (B) can also be used to solve various design problems wherein explicit information of the element connectivity matrix and nodal coordinates array are required, e.g., problems involving finite deformation \citep{saxena2013combined,kumar2016synthesis,kumar2017implementation,kumar2019computational,kumar2021topology}, linear elasticity-based problems, e.g., in \cite{sukumar2004conforming,tabarraei2006application,kumar2015topology,singh2020topology} and related references therein, etc. In addition, the TO approaches based on element design variables can be readily implemented and studied as the presented code provides uniform hexagonal tessellations.
 
 In summary, the primary goals of this paper are to provide a simple and efficient way to generate honeycomb tessellation\footnote{which is not trivial to generate}, TO with/without commonly used filtering techniques and optimality criteria approach, steps to include various design problems, and explicit expressions for the Wachspress shape functions and elemental stiffness matrix for the benefits of  new students to explore, learn and realize  TO with hexagonal elements in relatively less time. In addition, one can extend the presented code for advanced optimization problems involving stress and buckling constraints.

The remainder of the paper is organized as follows. Sec.~\ref{Sec:ProblemFormulation} briefly describes  the compliance minimization TO problem formulation with volume constraints, optimality criteria updating, sensitivity filtering and density filtering schemes for the sake of completeness. Sec.~\ref{Sec:ImplementationDetail} provides 
a novel,  compact and efficient way to generate element connectivity matrix and the corresponding nodal coordinates array for a honeycomb tessellations. In addition, finite element analysis, filtering and optimization procedures are also described, and  results of the Messerschmitt-Bolkow-Blohm  (MBB) beam are presented with and without filtering schemes. Further, to demonstrate distinctive features of the hexagonal elements, numerical examples for the beam design are presented, and results are compared with those obtained using quadrilateral elements. Sec.~\ref{Sec:SimpleExtension} presents the extensions of the code towards--different boundary conditions, multiloads situations, non-designs (passive) domains, and a Heaviside projection filtering scheme. In addition, directions for various other extensions are also reported. Lastly, conclusions are drawn in Sec.~\ref{Sec:Closure}.

	\section{Problem formulation}\label{Sec:ProblemFormulation}
\begin{figure}
	\centering
		\includegraphics[scale = 1.5]{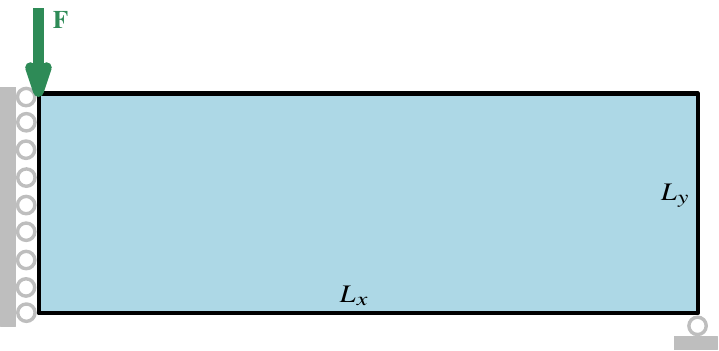}
	\caption{A symmetric MBB beam design domain with boundary conditions and an external force, $\mathbf{F}$}	\label{fig:MBBbeam}
\end{figure}

We consider the MBB beam design problem to demonstrate the presented Matlab TO code. Compliance of the beam is minimized for a given volume (resource) constraint.  A symmetric half design domain of the beam with the pertinent boundary conditions and external load $\mathbf{F}$ is depicted in Fig.~\ref{fig:MBBbeam}. $L_x$ and $L_y$ indicate dimensions in $x-$ and $y-$directions respectively herein and henceforth. 

The design domain is parameterized using \texttt{Nelem} hexagonal FEs represented via $\Omega_j^\text{H}|_{j=1,\,2,\,3,\,\cdots,\,\texttt{Nelem}}$. Each FE is assigned a design variable $\rho_j\in[0,\,1]$ that is constant within the element \citep{sigmund2007morphology}. The stiffness matrix of element $j$ is determined as 
\begin{equation}\label{EQ:elementStiff}
\mathbf{k}_j = E(\rho_j)\mathbf{k}_0 = \left(E_\text{min} + \rho_j^p\left(E_0 - E_\text{min}\right)\right)\mathbf{k}_0,
\end{equation}
where $E(\rho_j)$, Young's modulus of element $j$,  is evaluated using the modified Solid Isotropic Material with Penalization (SIMP) formulation \citep{sigmund2007morphology}. $E_0$  indicates the Young's modulus of a solid element ($\rho_j = 1$) and that of a void element ($\rho_j = 0$) is denoted via $E_\text{min}$. Material contrast i.e. $\frac{E_\text{min}}{E_0} = 10^{-9}$ is fixed to ensure nonsingular global stiffness matrix $\mathbf{K}$ \citep{sigmund200199}. $\mathbf{k}_0$ is the element stiffness matrix with $E(\rho_j) = 1$. The  SIMP parameter $p$ is set to 3 in this paper.

The following optimization problem is solved:
\begin{equation} \label{EQ:OPTI} 
\begin{rcases}
\begin{split}
&{\min_{\bm{\rho}}} \quad C({\bm{\rho}}) = \mathbf{u}^\top \mathbf{K}(\bm{\rho})\mathbf{u} = \sum_{j=1}^{\mathtt{Nelem}}\mathbf{u}_j^\top\mathbf{k}_j(\rho_j)\mathbf{u}_j\\
&\text{subjected to:}\\
&\bm{\lambda}:\,\,\mathbf{K} \mathbf{u} - \mathbf{F} = \mathbf{0}\\
&\Lambda:\,\,\text{g} = \frac{V(\bm{\rho})}{\mathtt{Nelem}\,V^*_f}-1 = \frac{\sum_{j=1}^\mathtt{Nelem}v_j\rho_j}{\mathtt{Nelem}\,V^*_f}-1 \le 0\\
&\quad\,\,\,\, \bm{0} \leq \bm{\rho} \leq \bm{1} \\
&\text{Data:} \quad \mathbf{F},\,V^*,\,V^*_f,\,v_j(=1),\,E_0,\,E_\text{min},\,p
\end{split}
\end{rcases},
\end{equation} 
where $C({\bm{\rho}})$ represents the compliance, $\mathbf{u}$ and $\mathbf{F}$ indicate the global displacement and force vectors respectively, and $\mathbf{u}_j$ is the displacement vector corresponding to element $j$. $V$ is the total material volume, and  $V^*_f$ is the permitted resource volume fraction of the design domain. $\bm{\rho}$, the design vector, is constituted via $\rho_j$. $\bm{\lambda}$ (vector) and $\Lambda$ (scalar) are the Lagrange multipliers corresponding to the state equilibrium equation and the volume constraint respectively. $\bm{\lambda = -2\mathbf{U}}$ can be found using the adjoint-variable method (see Appendix~\ref{append:sensanalysis}). Sensitivities of the objective with respect to $\rho_j$ using $\bm{\lambda}$ and \eqref{EQ:elementStiff} can be determined as (Appendix~\ref{append:sensanalysis})
\begin{equation}\label{EQ:Senstivity}
\frac{\partial C}{\partial \rho_j} = -\mathbf{u}_j^\top\frac{\partial \mathbf{k}_j}{\partial \rho_j} \mathbf{u}_j = -p (E_0-E_\text{min})\rho_j^{p-1}\mathbf{u}_j^\top \mathbf{k}_0\mathbf{u}_j.
\end{equation}
Likewise, derivatives of the volume constraint with respect to $\rho_j$ are determined as
\begin{equation}\label{EQ:volumesense}
\frac{\partial \text{g}}{\partial \rho_j} = \frac{v_j}{\mathtt{Nelem}\,V^*_f} = \frac{1}{\mathtt{Nelem}\,V^*_f},
\end{equation}
wherein $v_j =1$ is assumed.
The optimization problem \eqref{EQ:OPTI} is solved using the standard optimality criteria method wherein the design variables are updated per \cite{sigmund200199} given as
\begin{equation}
\rho_j^{k+1} =
\begin{lcases}
\begin{split}
&{(\rho_j^k)}_{-}\qquad \text{if}\,\mathcal{M}_j^k<{(\rho_j^k)}_{-},\\
&{(\rho_j^k)}_{+}\qquad \text{if}\,\mathcal{M}_j^k>{(\rho_j^k)}_{+},\\
&\mathcal{M}_j^k\qquad\,\,\,\, \text{otherwise},
\end{split}
\end{lcases}
\end{equation}
where ${(\rho_j^k)}_{-} = \max(0,\,\rho_j^k -m)$, ${(\rho_j^k)}_{+}= \min(1,\,\rho_j^k + m)$, and $m\in [0,\,1]$ indicates the move limit. \linebreak$\mathcal{M}_j^k = \rho_j^k{\left(\frac{-\frac{\partial C}{\partial\rho_j}}{\Lambda_k\frac{\partial \text{g}}{\partial \rho_j}}\right)}^{\frac{1}{2}}$, where $\Lambda_k$ is the value of $\Lambda$ at the $k^\text{th}$ iteration. The final value of $\Lambda$ is determined using the bisection algorithm \citep{sigmund200199}. 

We solve the optimization problem \eqref{EQ:OPTI} with and without filtering techniques.  Mesh-independent density filtering \citep{bruns2001topology,bourdin2001filters} and sensitivity filtering \citep{sigmund1997design,sigmund200199} are considered. As per  the density filtering \citep{bruns2001topology, xu2020bi}, the filtered density $\bar{\rho_j}$ of  element $j$ is evaluated as 
\begin{equation}\label{Eq:densityfilter}
\tilde{\rho_j} = \frac{\displaystyle\sum_{i\in n_j} v_i \rho_i w(\mathbf{x}_i)}{\displaystyle\sum_{i\in n_j} v_i w(\mathbf{x}_i)},
\end{equation}
where $v_i$ and $\rho_i$ are the volume and material density of the $i^\text{th}$ neighboring element respectively, and $n_j$ is the total number of neighboring elements of element~$j$ within a circle of radius $r_\text{fill}$. $w(\mathbf{x}_j)$, a linearly decaying weight function, is defined as \citep{bruns2001topology,bourdin2001filters,han2021topology}
\begin{equation}
w(\mathbf{x}_i) = \max \left(0,\frac{||\mathbf{x}_i - \mathbf{x}_j||}{r_\text{fill}}\right),
\end{equation}
$\mathbf{x}_i$ and $\mathbf{x}_j$ are the center coordinates of the $i^\text{th}$ and $j^\text{th}$ elements respectively, and $||.||$ defines a Euclidean distance. The chain rule is used to evaluate the final sensitivities of a function $f$ with respect to ${\rho_j}$. With the sensitivity filter, the filtered sensitivities are evaluated as \citep{sigmund1997design} 
\begin{equation}\label{Eq:sensfilter}
\frac{\overline{\partial C}}{\partial \rho_j} = \frac{\displaystyle\sum_{i\in n_j}w(\mathbf{x}_i)\rho_i \frac{\partial C}{\partial \rho_i}}{\max(\delta,\,\rho_j)\displaystyle\sum_{i\in n_j}w(\mathbf{x}_i)},
\end{equation}
where $\delta$, a small positive number, is set to $10^{-3}$ for avoiding division by zero \citep{andreassen2011efficient}.

	\section{Implementation detail}\label{Sec:ImplementationDetail}
In this section, MATLAB implementation of \texttt{HoneyTop90} is presented in detail. We first provide a novel, efficient and simple code to generate honeycomb elements connectivity matrix and corresponding nodal coordinates array in just 9~(13) lines.  Thereafter, finite element analysis, filtering and optimization  are described for the presented \texttt{HoneyTop90} TO code. The MBB beam is optimized herein to demonstrate \texttt{HoneyTop90}. 
\begin{figure}[h!]
	\centering
	\includegraphics[scale = 1]{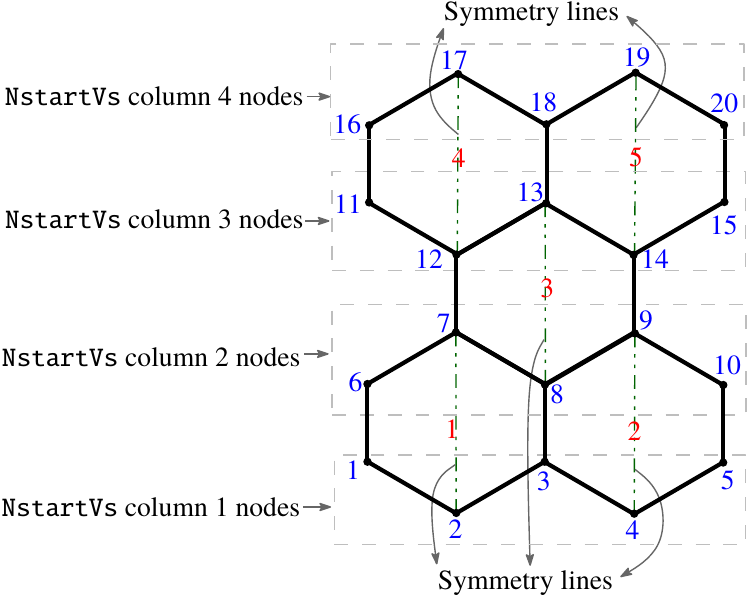}
	\caption{A schematic diagram for element and node numbering steps. Texts in blue and red indicate node and element numbers respectively, and henceforth the same colors  are used to indicate nodes and elements. \texttt{NstartVs} is determined on the line~6 of the \texttt{HoneyTop90} code (Appendix~\ref{app:MATLABcode}).}	\label{fig:hexnodelement}
\end{figure}
\subsection{Element connectivity and nodal coordinates matrices (lines 5-18)} \label{SubSec:ElementConnNodalConn}
Let $\mathtt{HNex}$ and $\mathtt{HNey}$ be the number of hexagonal elements in $x-$ and $y-$directions respectively. Each element consists of six nodes, and each node possesses two degrees of freedom (DOFs). For element~$i$, DOFs $2i-1$ and $2i$ correspond respectively to the displacement in $x-$ and $y-$directions. In the provided code (Appendix~\ref{app:MATLABcode}), the element DOFs matrix \texttt{HoneyDOFs} is generated on lines 6-10, and the corresponding nodal coordinate matrix \texttt{HoneyNCO} is generated on lines 11-14. When $\mathtt{HNey}$ is  an even number, corresponding \texttt{HoneyDOFs} and \texttt{HoneyNCO} are updated on lines 15-18. The $i^\text{th}$ row of \texttt{HoneyDOFs} gives DOFs corresponding to element $i$, whereas that of \texttt{HoneyNCO} contains $x-$ and $y-$coordinates of node $i$.

\begin{figure}
	\centering
	\includegraphics[scale = 1]{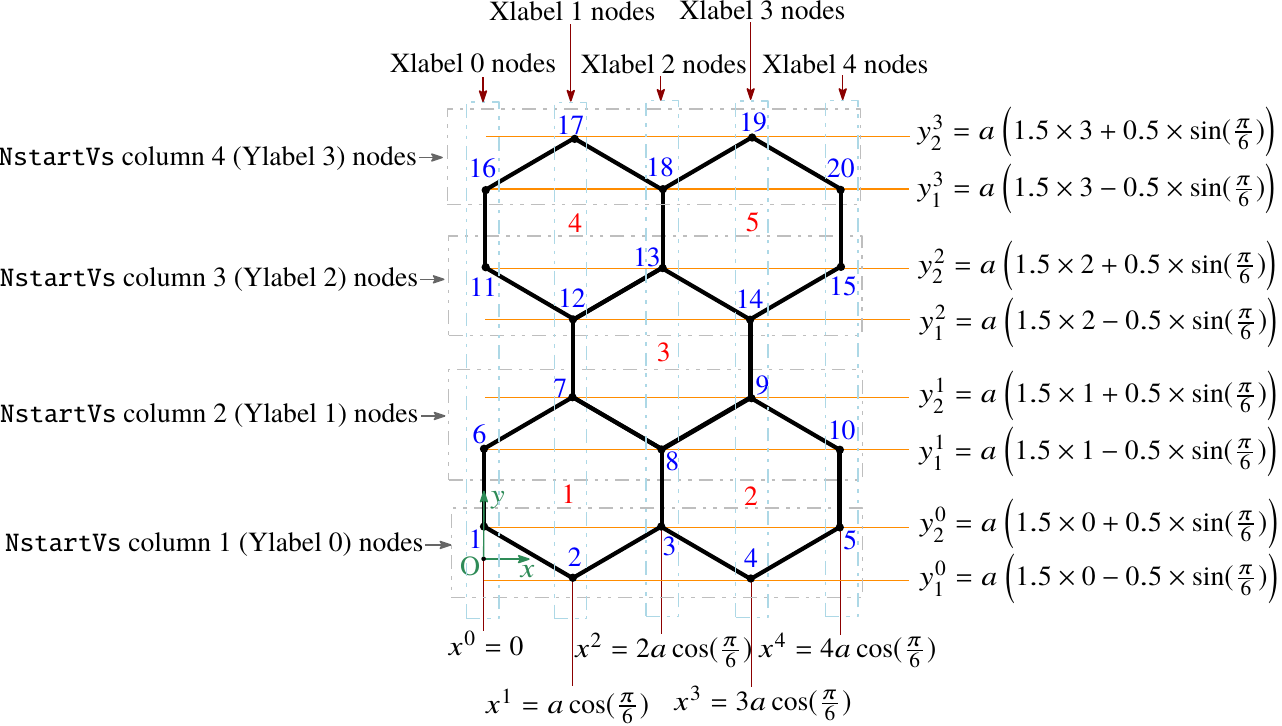}
	\caption{Finding the nodal coordinates of a honeycomb tessellation. $\mathbf{O}$ is the origin.}	\label{fig:hexnodalcoordinates}
\end{figure}

Columns of the matrix \texttt{NstartVs} indicate node numbers of FEs starting from the bottom to top rows (Fig.~\ref{fig:hexnodelement}). The first $x-$DOFs of both symmetrical half quadrilaterals (see Fig.~\ref{fig:hexnodelement}) of each hexagonal FE are stored in the matrix \texttt{DOFstartVs}. DOFs of all such quadrilaterals are recorded in the matrix \texttt{NodeDOFs}. The redundant rows of the matrix \texttt{NodeDOFs} are removed using \texttt{setdiff} MATLAB function, and the remaining DOFs are noted in the matrix \texttt{ActualDOFs}. The final honeycomb DOFs connectivity matrix  \texttt{HoneyDOFs} is obtained from the matrix \texttt{ActualDOFs} and recorded on line~10. 

\begin{figure}[h!]
	\centering
	\begin{subfigure}{0.3\textwidth}
		\centering
		\includegraphics[scale = 0.35]{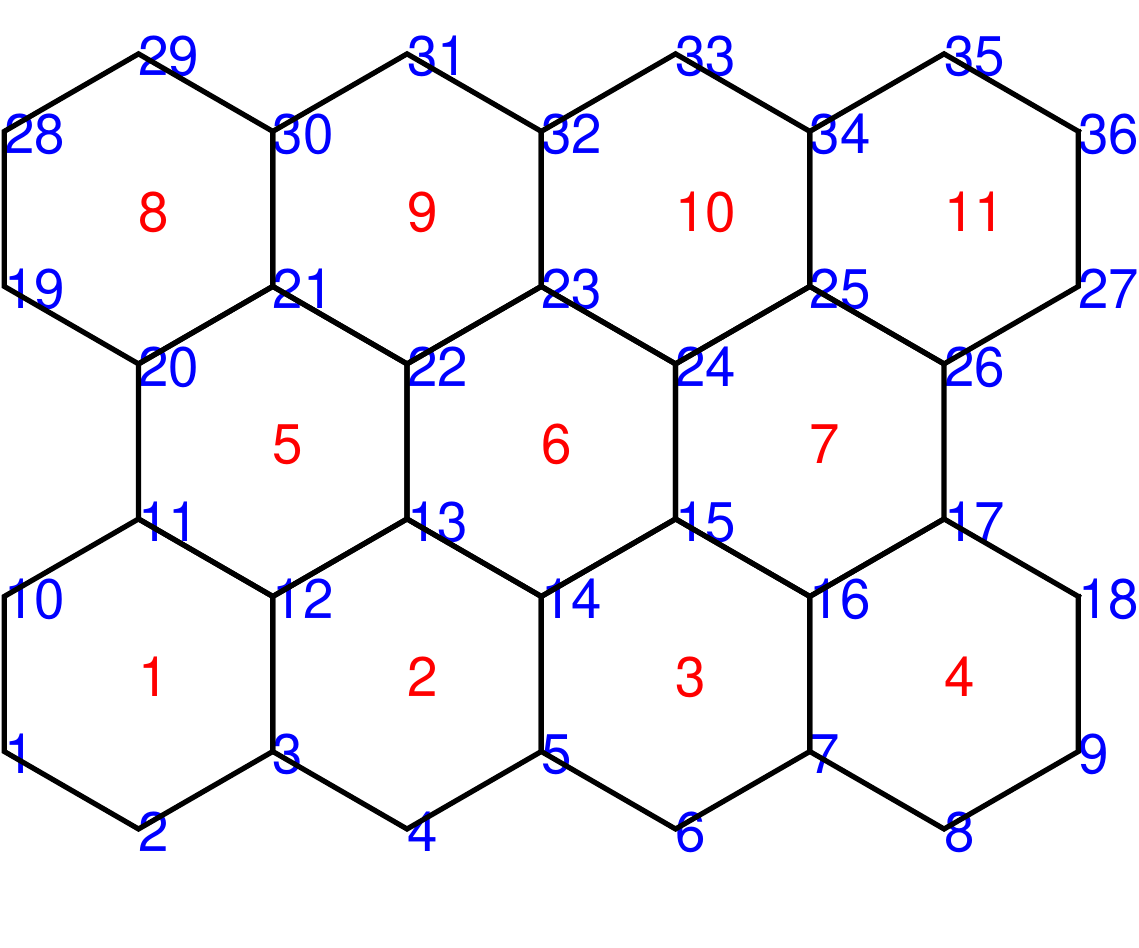}
		\caption{$\mathtt{HNex} = 4,\,\mathtt{HNey} =3$}
		\label{fig:hex_odd}
	\end{subfigure}
	\quad
	\begin{subfigure}{0.3\textwidth}
		\centering
		\includegraphics[scale = 0.35]{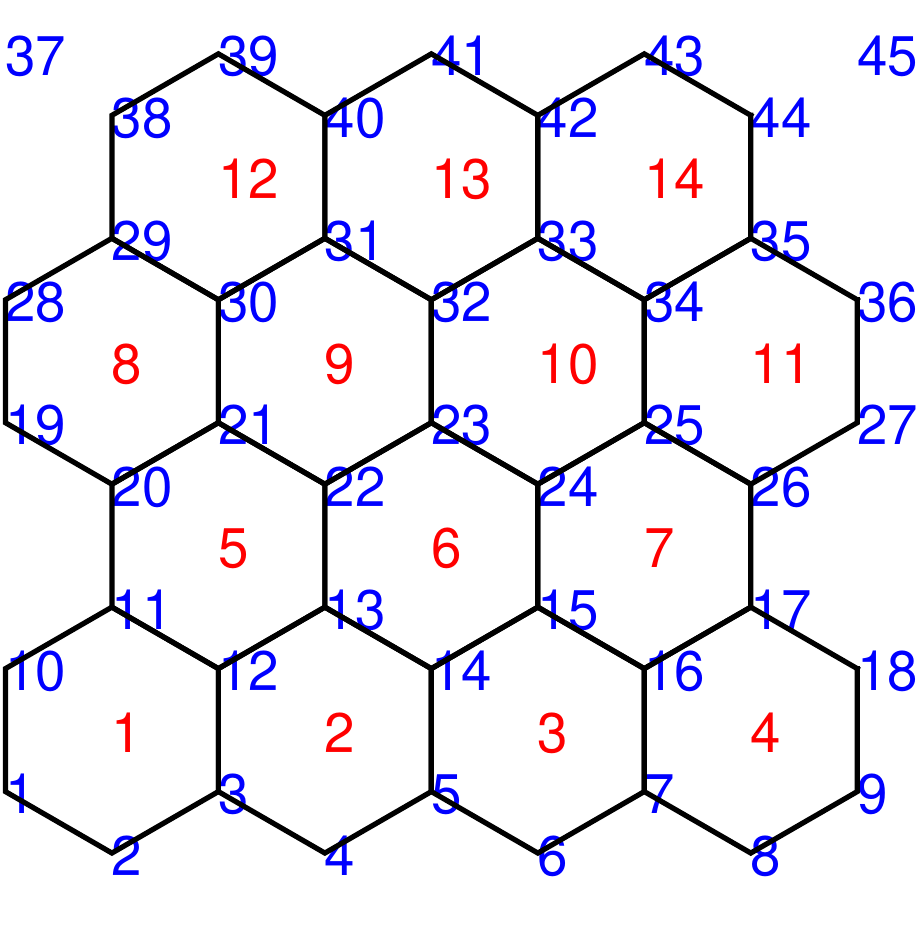}
		\caption{$\mathtt{HNex} = 4,\,\mathtt{HNey} =4$.  37 and 45 are the hanging nodes.}
		\label{fig:hex_even}
	\end{subfigure}
	\quad
	\begin{subfigure}{0.3\textwidth}
		\centering
		\includegraphics[scale = 0.35]{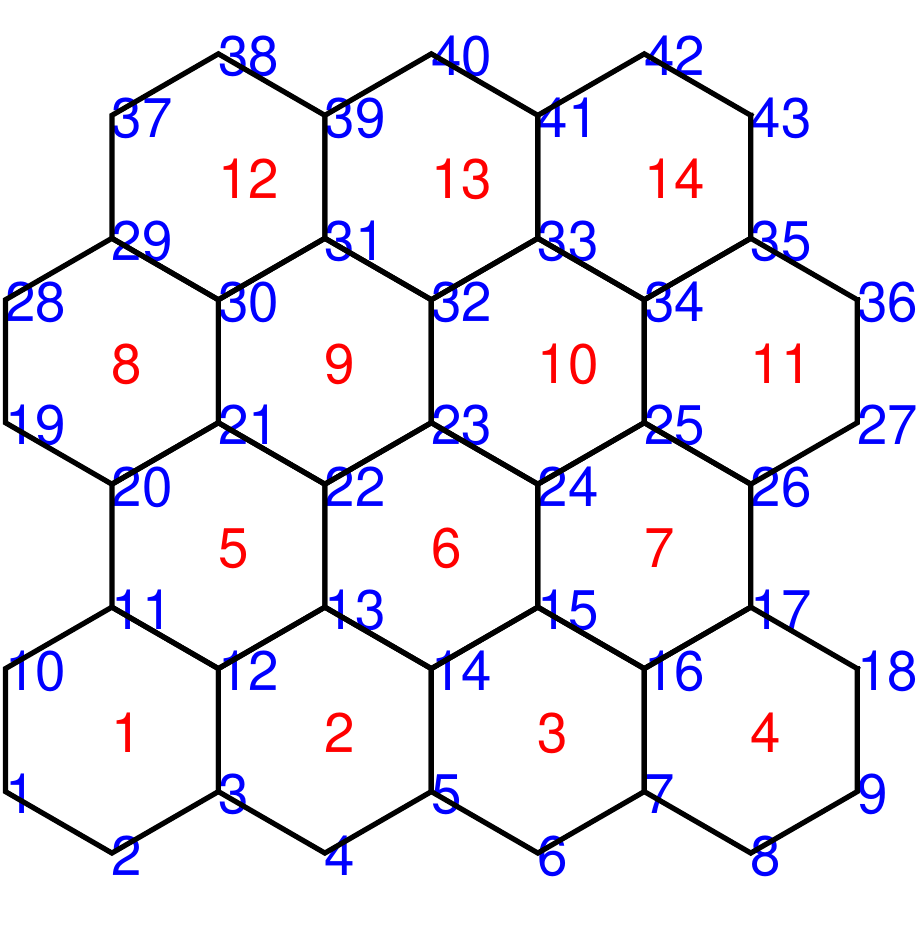}
		\caption{$\mathtt{HNex} = 4,\,\mathtt{HNey} =4$. Hanging nodes are removed (lines 15-18) and node numbering is updated.}
		\label{fig:hex_even_updated}
	\end{subfigure}
	\caption{Different honeycomb tessellations. Hanging nodes are detected in (\subref{fig:hex_even}), which  are removed, and the  updated node numbers with respective elements are plotted in (\subref{fig:hex_even_updated}).}	\label{fig:Hexdifferentcases}
\end{figure}

Coordinates of vertex $m|_{m= 0,\,1,\,\cdots,\,6}$ of a hexagonal element with centroid (a local coordinate system) at the origin can be written as $\left(a\cos(\frac{(2m-1)\pi}{6},\,a\sin(\frac{(2m-1)\pi}{6})\right)$, where $a$ is the length of an edge, which can have different values as desired. When the origin is shifted to ($-a\cos(\frac{\pi}{6}),\,-0.75a$) with respect to the local coordinate system of element 1 (Fig.~\ref{fig:hexnodalcoordinates}), the $y-$coordinates of the vertices for a honeycomb tessellation  can be written as (Fig.~\ref{fig:hexnodalcoordinates})
\begin{equation}\label{Eq:ycoordinates}
	\begin{split}
		y_k^l = a\left(1.5l + \frac{(-1)^k}{2}\sin\left(\frac{\pi}{6}\right)\right), \\
		\text{with}\, k = 1,\,2;\,l = 0,\,1,\,\cdots,\,\texttt{HNey}
	\end{split}
\end{equation}
Likewise, the  $x-$coordinates can be written as (Fig.~\ref{fig:hexnodalcoordinates})
\begin{equation}\label{Eq:xcoordinates}
	x^n = na\cos\left(\frac{\pi}{6}\right),
	\text{with}\, n = 0,\,1,\,\cdots,\,\texttt{2HNex}
\end{equation}

In view of \eqref{Eq:ycoordinates}, the $y-$coordinates of the nodes for a general honeycomb tessellation, i.e.,  corresponding to the matrix \texttt{HoneyDOFs}, are determined on lines 11-13 and stored in the vector \texttt{Ncyf}.  With the $x-$coordinates \eqref{Eq:xcoordinates}, the matrix \texttt{HoneyNCO} records the nodal coordinates of the meshgrid on line~14, wherein  $x-$ and $y-$coordinates are kept in the first and second columns respectively. In the provided code (Appendix~\ref{app:MATLABcode}), $a$ is taken $1/\sqrt{3}$ (lines 14 and 43). For the desired edge length $\kappa$, the user should replace $1$ with $\kappa$ on lines 14 and 43 (centroid determination) of \texttt{HoneyTop90} code and may
have to accordingly determine the shape functions and the elemental stiffness matrix.

\begin{figure}[h!]
	\centering
	\begin{subfigure}{0.49\textwidth}
		\centering
		\includegraphics[scale = 0.60]{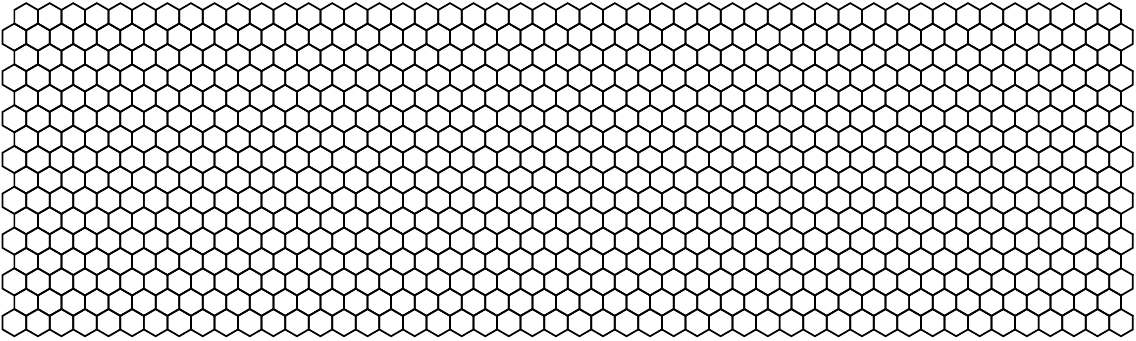}
		\caption{$48\times16$ hexagonal elements}
		\label{fig:MBB90_30}
	\end{subfigure}
	\begin{subfigure}{0.49\textwidth}
		\centering
		\includegraphics[scale = 0.600]{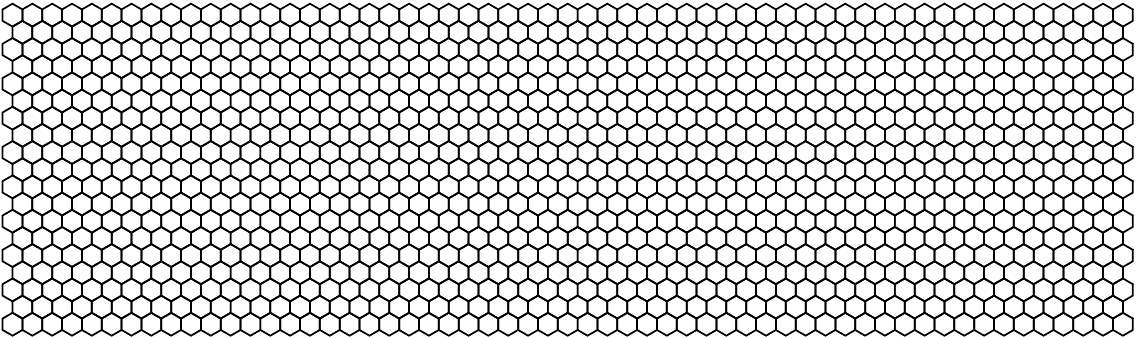}
		\caption{$57\times19$ hexagonal elements}
		\label{fig:MBB145_45}
	\end{subfigure}
	\caption{Honeycomb tessellations for the MBB beam.}	\label{fig:MBBdiscretization}
\end{figure}

When $\mathtt{HNey}$ is  an even number, hanging nodes are observed (Fig.~\ref{fig:hex_even}) with the above presented steps. Hanging nodes are removed (lines 15-18), and the connectivity DOFs matrix \texttt{HoneyDOFs} is updated on line~16 accordingly. Likewise, \texttt{HoneyNCO} is updated on line 17 by removing the hanging nodes (Fig.~\ref{fig:hex_even}).   One notices that  the DOFs and nodal coordinates matrices are determined primarily by using  \texttt{reshape} and \texttt{repmat} MATLAB functions. The former rearranges the given matrix for the specified number of rows and columns consistently, whereas the latter duplicates the matrix for the assigned number of times along the $x-$ and $y-$directions. The obtained element connectivity matrices for Fig.~\ref{fig:hex_odd} and Fig.~\ref{fig:hex_even_updated} are noted below in the matrices $\texttt{HoneyDOFs}_{4\times 3}$ \eqref{Eq:Hex4_3} and $\texttt{HoneyDOFs}_{4\times 4}$ \eqref{Eq:Hex4_4} respectively. Entries in row $i|_{i = 1,\,2\,\cdots,\,\mathtt{Nelem}}$ of these matrices indicate DOFs corresponding to element $i$. One can determine honeycomb element connectivity matrix $\mathtt{HoneyElem}$  from the \texttt{HoneyDOFs} as,
\begin{lstlisting}[basicstyle=\scriptsize\ttfamily,breaklines=true,numbers=none]
	HoneyElem = HoneyDOFs(:,2:2:end)/2.
\end{lstlisting}
The $r^\text{th}$ row of the matrix \texttt{HoneyElem} contains nodes  in the counter-clockwise sense that constitute element $r$.  \texttt{HoneyNCO} and \texttt{HoneyElem} provide an important set of ingredients for performing finite element analysis using hexagonal FEs. The following code can be used to plot and visualize the honeycomb tessellation generated by the steps mentioned above (lines 6-18).
\begin{lstlisting}[numbers = none]
	% Code to plot honeycomb tessellation
	H = figure(1);set(H,'color','w');
	X = reshape(HoneyNCO(HoneycombElem',1),6,Nelem);
	Y = reshape(HoneyNCO(HoneycombElem',2),6,Nelem);
	patch(X, Y,'w','EdgeColor','k');axis off equal;
\end{lstlisting}
The MBB beam (Fig.~\ref{fig:MBBbeam}) is meshed using $48\times16$ and $57\times19$ hexagonal elements\footnote{Coarse FEs are used for clear visibility.} to illustrate the discretization part of $\texttt{HoneyTop90}$ (Fig.~\ref{fig:MBBdiscretization}). Note that instead of removing hanging nodes as described above, one can also use those nodes to close the rectangular domain by forming triangular and quadrilateral elements at the boundaries. However, the computational cost of the entire optimization process can increase. This may happen  partly due to the requirement of different bookkeeping to store the element connectivity and  stiffness matrices of hexagonal, quadrilateral and triangular elements and partly because,  with different element types, nodal design variable-based TO methods need to be employed instead of elemental design variable-based TO approaches.

Next, the computation time taken by the code to generate \texttt{HoneyDOFs} and \texttt{HoneyNCO} for different meshgrids is noted in Table~\ref{Tab:Computationtime}. It can be observed that the DOFs connectivity matrix and corresponding nodal coordinates array for $3000\times 1000$ FEs can be generated within a fraction of a second.

\begin{equation}\label{Eq:Hex4_3}
	\texttt{HoneyDOFs}_{4\times 3} = 
	\begin{bmatrix}
		23 &   24   & 21  &  22   & 19  &  20  &   1  &   2   &  3  &   4   &  5   &  6 \\
		27  &  28 &   25 &   26  &  23   & 24  &   5 &    6 &    7 &    8  &   9   & 10\\
		31  &  32  &  29   & 30 &   27  &  28  &   9  &  10 &   11   & 12 &   13   & 14\\
		\vdots &  \vdots  &  \vdots &  \vdots   &\vdots & \vdots  & \vdots   &\vdots  &  \vdots   & \vdots  &  \vdots &   \vdots\\
		71  &  72  &  69  &  70 &   67  &  68   & 49   & 50  &  51   & 52  &  53  &  54 
	\end{bmatrix}
\end{equation}
\begin{equation}\label{Eq:Hex4_4}
	\texttt{HoneyDOFs}_{4\times 4} = 
	\begin{bmatrix}
		23 &   24   & 21  &  22   & 19  &  20  &   1  &   2   &  3  &   4   &  5   &  6 \\
		27  &  28 &   25 &   26  &  23   & 24  &   5 &    6 &    7 &    8  &   9   & 10\\
		31  &  32  &  29   & 30 &   27  &  28  &   9  &  10 &   11   & 12 &   13   & 14\\
		\vdots &  \vdots  &  \vdots &  \vdots   &\vdots & \vdots  & \vdots   &\vdots  &  \vdots   & \vdots  &  \vdots &   \vdots\\
		87  &  88 &   85 &   86    &83   & 84  &  65   & 66 &   67   & 68   & 69  &  70
	\end{bmatrix}
\end{equation}
\begin{table}[h!]
	\centering
	\caption{Computation time required to generate different mesh sizes. A 64-bit operating system machine with 8.0 GB RAM, Intel(R), Core(TM)~i5-8265U~CPU 1.60~GHz is used.}\label{Tab:Computationtime}
	\begin{tabular}{c|c|c}
		\hline \hline   
		\multicolumn{1}{c|}{\multirow{2}{*}{Mesh Size ($\mathtt{HNex}\times \mathtt{HNey}$)}} & \multicolumn{2}{c}{Computation time (s)} \\ \cline{2-3} 
		\multicolumn{1}{c|}{}                         & \texttt{HoneyDOFs}            & \texttt{HoneyNCO}            \\ \hline \rule{0pt}{3ex}
		$100 \times 50$                    &     0.031                &           0.0022           \\ \hline \rule{0pt}{3ex}
		$250 \times 125$                    &      0.037              &          0.0025              \\ \hline \rule{0pt}{3ex}
		$500 \times 250$                    &         0.066           &           0.011            \\ \hline \rule{0pt}{3ex}
		$1000 \times 500$                   &        0.156             &      0.032                \\ \hline \rule{0pt}{3ex}
		$2000 \times 1000$                  &       0.553                  &     0.135                \\ \hline\rule{0pt}{3ex}
		$3000 \times 1000$                  &       0.753                 &     0.178                \\ \hline \hline
	\end{tabular}
\end{table}
\subsection{Finite element analysis (lines 20-38 and lines 60-62)}\label{SubSec:FEA}

\texttt{Nelem} and $\mathtt{Nnode}$  (the total number of nodes) for the current meshgrid are determined using $\mathtt{size}$  and $\mathtt{deal}$ MATLAB functions on line 19. Alternatively, one can also use the following codes to determine them:

\begin{lstlisting}[basicstyle=\scriptsize\ttfamily,breaklines=true,numbers=none]
	Nelem = HNex*ceil(HNey/2) + (HNex-1)*floor(HNey/2);
	if(mod(HNey,2)==0)
	Nnode = (2*HNex+1)*(HNey+1)-2; % When HNey is an even	
	else
	Nnode = (2*HNex+1)*(HNey+1);  % When HNey is an odd
	end
\end{lstlisting}

The total DOFs is listed in \texttt{alldof} on line 23. The material Young's modulus $E_0$ and that of a void element $E_\text{min}$ are respectively denoted by $\mathtt{E0}$ and $\mathtt{E\text{min}}$. The Poisson's ratio $\nu = 0.29$ is considered. The elemental stiffness matrix~\texttt{ Ke} is mentioned on lines 27-38 that is evaluated using the Wachspress shape functions \citep{wachspress1975rational} with plane stress assumptions. \cite{talischi2009honeycomb}  also employ   these shape functions (see Appendix~\ref{appsec:Wachspress_shape_functions} and Appendix~\ref{appsec:Numerical_quadrature_points}). However,  a method to generate honeycomb tessellation which is not so straightforward, steps to include various problems, explicit expressions for the shape functions  and a MATLAB code to learn and extend the method are not presented in \cite{talischi2009honeycomb}.

The global stiffness matrix $\mathtt{K}$ is evaluated  on line~61 using the \texttt{sparse} function.  The rows and columns of the matrix \texttt{HoneyDOFs} are recorded in vector \texttt{iK} and \texttt{jk} respectively \citep{andreassen2011efficient}.  Boundary conditions of the given design description are recorded in the vector \texttt{fixeddofs} on line~22, and the applied external force is noted in the vector \texttt{F} on line~20. The vectors  \texttt{fixeddofs} and \texttt{F} can be modified based on the different problem settings. The displacement vector  \texttt{U} is determined on line~62.

\subsection{Filtering, Optimization and Results Printing (lines 39-52 and lines 63-89)}\label{SubSec:Filtering}
We provide three filtering cases for a problem---sensitivity filtering~(\texttt{ft = 1}), density filtering (\texttt{ft=2}) and null (no) filtering (\texttt{ft =0}). The latter is included to demonstrate the characteristics of hexagonal elements in view of checkerboard patterns and point connections. For generating the required filter matrices (\texttt{DD} and \texttt{HHs}, lines 44-52), we need the coordinates of centroids of the hexagonal elements. For that, the center coordinates array \texttt{ct} is determined on lines~40-43. The $i^\text{th}$ row of \texttt{ct} gives  $x-$ and $y-$coordinates of the centroid of element $i$. \texttt{ct} matrix is used to evaluate filtering parameter and also, is employed to plot the intermediate results using the \texttt{scatter} MATLAB function. 

The neighboring elements of each element for the given filter radius \texttt{rfill} are determined on line~47. They are stored in the matrix \texttt{DD} whose  first, second and third entries indicate the neighborhood FE index, the selected element index and center-to-center distance between them, respectively. The filtering matrix \texttt{HHs} is evaluated on line~52 using the \texttt{spdiags} function \citep{talischi2012polytop}. S = \texttt{spdiags}(Bin,\,d,\,m,\,n) creates an m-by-n sparse matrix S by taking the columns of Bin and placing them along the diagonals specified by d. 

The given volume fraction \texttt{volfrac} is used herein to set the initial guess of TO. The design vector  $\bm{\rho}$ is denoted by \texttt{x} in the code.  The  physical material density vector is represented by \texttt{xPhys}. With either $\mathtt{ft=0}$ or $\mathtt{ft=1}$,  \texttt{xPhys} is equal to the design vector, whereas with $\mathtt{ft=2}$, \texttt{xPhys} is represented via the filtered density $\tilde{\bm{\rho}}$ (line~84). The optimization updation is performed on line~81 as per \cite{ferrari2020new}. The objective, i.e., compliance of the design is obtained on line~65. The mechanical equilibrium equation is solved to determine displacement vector \texttt{U} on line~62 using the \texttt{decomposition} function with lower triangle part of \texttt{K}. \texttt{decomposition} function provides efficient Cholesky decomposition of the global stiffness matrix, and thus, solving state equations becomes computationally cheap. The sensitivities of the objective are evaluated and stored in the vector \texttt{dc} on line~66. The vector \texttt{dc} is updated as per different $\mathtt{ft}$ values. Volume constraint sensitivities are recorded in the vector \texttt{dv} and updated within the loop based on chosen filtering technique.  For plotting and visualizing the intermediate results evolution, we use \texttt{scatter}  function as on line~89.  The plotting function uses the centroid coordinates in conjunction with the intermediate physical design vector \texttt{xPhys}. Alternatively, one can also use 
\begin{lstlisting}[basicstyle=\scriptsize\ttfamily,breaklines=true,numbers=none]
	colormap('gray');
	X = reshape(HoneyNCO(HoneyElem',1),6,Nelem);
	Y = reshape(HoneyNCO(HoneyElem',2),6,Nelem);
	patch(X, Y, [1-xPhys],'EdgeColor','k');
	axis equal off; pause(1e-6);
\end{lstlisting}
to plot the intermediate results.
\subsection{MBB beam optimized results}\label{SubSec:MBBOptimizedResult}
\texttt{HoneyTop90} MATLAB code is provided in Appendix~\ref{app:MATLABcode}. The code is called as
\begin{lstlisting}[numbers = none]
	HoneyTop90(HNex,HNey,volfrac,penal,rfill,ft);
\end{lstlisting}
to find the optimized beam designs for the domain and boundary conditions shown in Fig.~\ref{fig:MBBbeam}. The results are obtained with and without filtering schemes, i.e., \texttt{ft = 0,\,1,\,2} are used. \texttt{volfrac}, the permitted volume fraction, is set to 0.5. The SIMP parameter denoted by \texttt{penal} is set to 3. Three mesh sizes with $60 \times 20$, $150 \times 50$ and $300 \times 100$ FEs are considered. The filter radius \texttt{rfill} is set to 0.03 times the length of the beam domain, i.e., $1.8\sqrt{3}$, $4.5\sqrt{3}$ and $9\sqrt{3}$ 
for the $60 \times 20$, $150 \times 50$ and $300 \times 100$ FEs respectively.
\begin{figure}[h!]
	\centering
	\begin{subfigure}{0.30\textwidth}
		\centering
		\includegraphics[scale = 0.38]{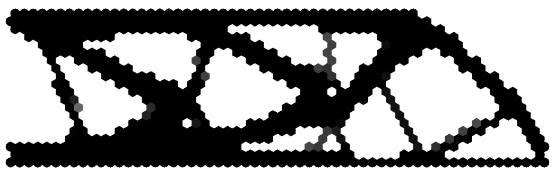}
		\caption{$C = 298.70$}
		\label{fig:Hex6020ft0}
	\end{subfigure}
	\begin{subfigure}{0.30\textwidth}
		\centering
		\includegraphics[scale = 0.380]{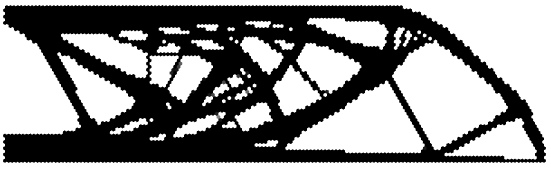}
		\caption{$C=290.47$}
		\label{fig:Hex15050ft0}
	\end{subfigure}
	\begin{subfigure}{0.30\textwidth}
		\centering
		\includegraphics[scale = 0.38]{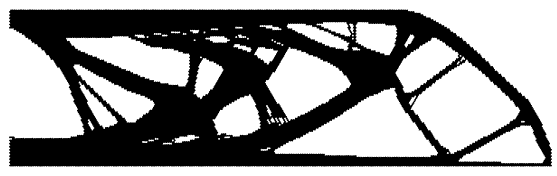}
		\caption{$C=282.49$}
		\label{fig:Hex300100ft0}
	\end{subfigure}
	\begin{subfigure}{0.30\textwidth}
		\centering
		\includegraphics[scale = 0.38]{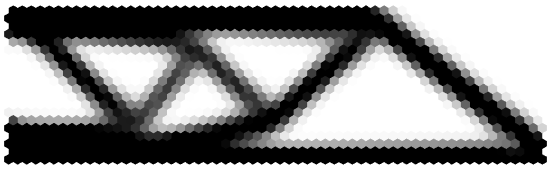}
		\caption{$C = 307.93$}
		\label{fig:Hex6020ft1}
	\end{subfigure}
	\begin{subfigure}{0.30\textwidth}
		\centering
		\includegraphics[scale = 0.38]{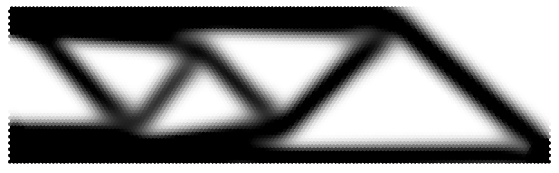}
		\caption{$C=309.54$}
		\label{fig:Hex15050ft1}
	\end{subfigure}
	\begin{subfigure}{0.30\textwidth}
		\centering
		\includegraphics[scale = 0.380]{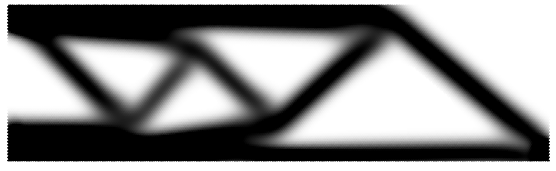}
		\caption{$C=302.33$}
		\label{fig:Hex300100ft1}
	\end{subfigure}
	\begin{subfigure}{0.30\textwidth}
		\centering
		\includegraphics[scale = 0.38]{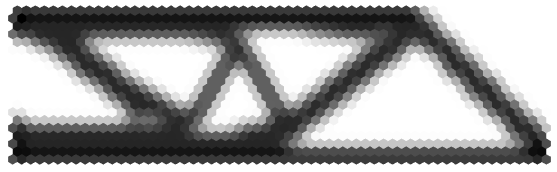}
		\caption{$C =337.44$}
		\label{fig:Hex6020ft2}
	\end{subfigure}
	\begin{subfigure}{0.30\textwidth}
		\centering
		\includegraphics[scale = 0.380]{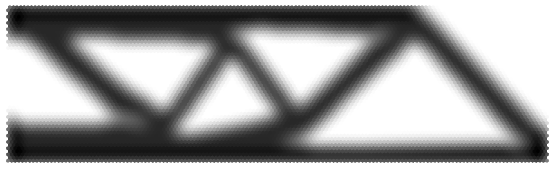}
		\caption{$C= 358.63$}
		\label{fig:Hex15050ft2}
	\end{subfigure}
	\begin{subfigure}{0.30\textwidth}
		\centering
		\includegraphics[scale = 0.380]{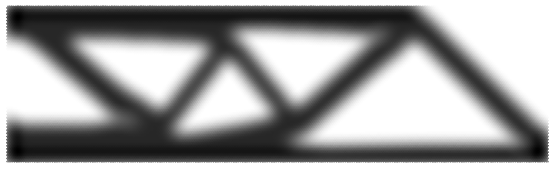}
		\caption{$C=367.64$}
		\label{fig:Hex300100ft2}
	\end{subfigure}
	\caption{The optimized MBB beam designs are displayed. The results with corresponding compliance $C$ depicted in  the first (\subref{fig:Hex6020ft0}-\subref{fig:Hex300100ft0}), second (\subref{fig:Hex6020ft1}-\subref{fig:Hex300100ft1}) and third~(\subref{fig:Hex6020ft2}-\subref{fig:Hex300100ft2}) rows are obtained with no filtering, sensitivity filtering and density filtering respectively.}	\label{fig:MBBresults}
\end{figure}

The optimized results obtained with different filtering approaches are displayed in Fig.~\ref{fig:MBBresults}. The results  with \texttt{ft=0} (Fig.~\ref{fig:Hex6020ft0}-\ref{fig:Hex300100ft0}) are free from checkerboard patterns and are the best performing ones. They are however mesh dependent  and contain thin members as expected. Therefore, to circumvent such features one can use either filtering or length scale constraints, e.g., perimeter constraint \citep{haber1996new}. Results displayed  in the second (Fig.~\ref{fig:Hex6020ft1}-\ref{fig:Hex300100ft1}) and third (Fig.~\ref{fig:Hex6020ft2}-\ref{fig:Hex300100ft2}) rows are obtained with sensitivity filtering (\texttt{ft=1})  and density filtering (\texttt{ft=2}) respectively. These designs do not have checkerboard patterns and also, they are mesh independent, i.e., they have same topology irrespective of the mesh sizes employed. In addition, obtained topologies with sensitivity filtering and density filtering are same herein but that may not be the case always since both filters work on different principles as noted in Sec.~\ref{Sec:ProblemFormulation}. Figure 8 depicts the objective convergence plots for the MBB beam design for 200 OC iterations. One can note that objective history plots are stable and have converging nature.  

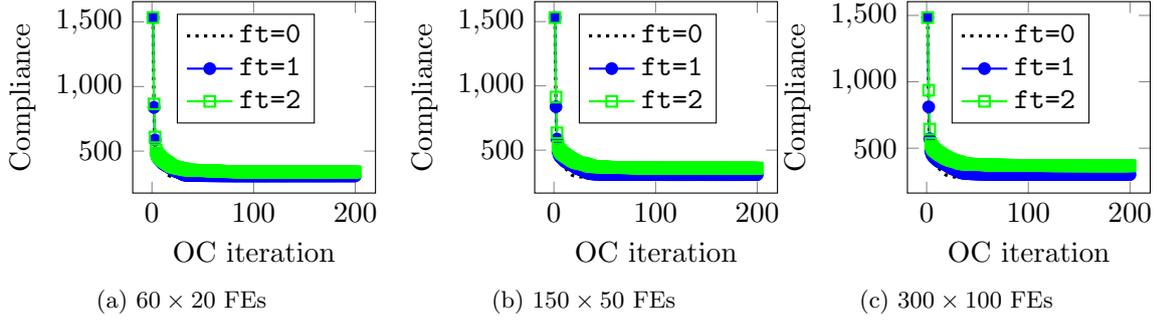
\begin{figure}[h!]
	\begin{subfigure}[t]{0.30\textwidth}
		\centering
		\begin{tikzpicture} 	
			\pgfplotsset{compat =1.9}
			\begin{axis}[
				width = 1\textwidth,
				xlabel= OC iteration,
				ylabel= Compliance,
				legend style={at={(0.75,0.65)},anchor=east}]
				\pgfplotstableread{60by20ft0.txt}\mydata;
				\addplot[smooth,black, mark size=2pt,style={very thick}, dotted]
				table {\mydata};
				\addlegendentry{\texttt{ft=0}}
				\pgfplotstableread{60by20ft1.txt}\mydata;
				\addplot[smooth,blue,mark = *, mark size=2pt,style= thick]
				table {\mydata};
				\addlegendentry{\texttt{ft=1}}
				\pgfplotstableread{60by20ft2.txt}\mydata;
				\addplot[smooth,green,mark = square,mark size=2pt,style= thick]
				table {\mydata};
				\addlegendentry{\texttt{ft=2}}
			\end{axis}
		\end{tikzpicture}
		\caption{$60\times 20$ FEs}
		\label{fig:60by20}
	\end{subfigure}
	\quad
	\begin{subfigure}[t]{0.30\textwidth}
		\centering
		\begin{tikzpicture} 	
			\pgfplotsset{compat =1.9}
			\begin{axis}[
				width = 1\textwidth,
				xlabel= OC iteration,
				ylabel= Compliance,
				legend style={at={(0.75,0.65)},anchor=east}]
				\pgfplotstableread{150by50ft0.txt}\mydata;
				\addplot[smooth,black, mark size=2pt,style={very thick}, dotted]
				table {\mydata};
				\addlegendentry{\texttt{ft=0}}
				\pgfplotstableread{150by50ft1.txt}\mydata;
				\addplot[smooth,blue,mark = *, mark size=2pt,style= thick]
				table {\mydata};
				\addlegendentry{\texttt{ft=1}}
				\pgfplotstableread{150by50ft2.txt}\mydata;
				\addplot[smooth,green,mark = square,mark size=2pt,style= thick]
				table {\mydata};
				\addlegendentry{\texttt{ft=2}}
			\end{axis}
		\end{tikzpicture}
		\caption{$150\times 50$ FEs}
		\label{fig:150by50}
	\end{subfigure}
	\begin{subfigure}[t]{0.30\textwidth}
		\centering
		\begin{tikzpicture} 	
			\pgfplotsset{compat =1.9}
			\begin{axis}[
				width = 1\textwidth,
				xlabel= OC iteration,
				ylabel= Compliance,
				legend style={at={(0.75,0.65)},anchor=east}]
				\pgfplotstableread{300by100ft0.txt}\mydata;
				\addplot[smooth,black, mark size=2pt,style={very thick}, dotted]
				table {\mydata};
				\addlegendentry{\texttt{ft=0}}
				\pgfplotstableread{300by100ft1.txt}\mydata;
				\addplot[smooth,blue,mark = *, mark size=2pt,style= thick]
				table {\mydata};
				\addlegendentry{\texttt{ft=1}}
				\pgfplotstableread{300by100ft2.txt}\mydata;
				\addplot[smooth,green,mark = square,mark size=2pt,style= thick]
				table {\mydata};
				\addlegendentry{\texttt{ft=2}}
			\end{axis}
		\end{tikzpicture}
		\caption{$300 \times 100$ FEs}
		\label{fig:300by100}
	\end{subfigure}
	\caption*{Figure 8: Objective convergence plots of the MBB beam for 200 optimality criteria (OC) iterations.}
	\label{fig:InverterOConvergence}
\end{figure}

\subsubsection{Checkerboard-patterns-free designs}
	In this section, the results obtained with hexagonal elements are compared with the corresponding quadrilateral FEs. The 88-line code, $\texttt{top88}$ \citep{andreassen2011efficient}, is employed to generate the results with quadrilateral FEs. Filtering techniques are not used. The volume fraction is set to 0.5 and the SIMP parameter $p=3$ is used.
\begin{figure}[h!]
	\centering
	\begin{subfigure}{0.45\textwidth}
		\centering
		\includegraphics[scale = 0.20]{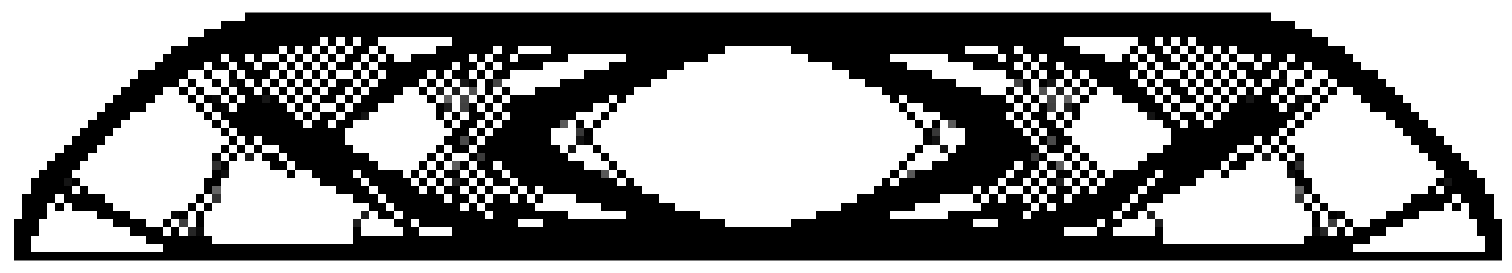}
		\caption{}
		\label{fig:quad90by30}
	\end{subfigure}
	\begin{subfigure}{0.45\textwidth}
		\centering
		\includegraphics[scale = 0.20]{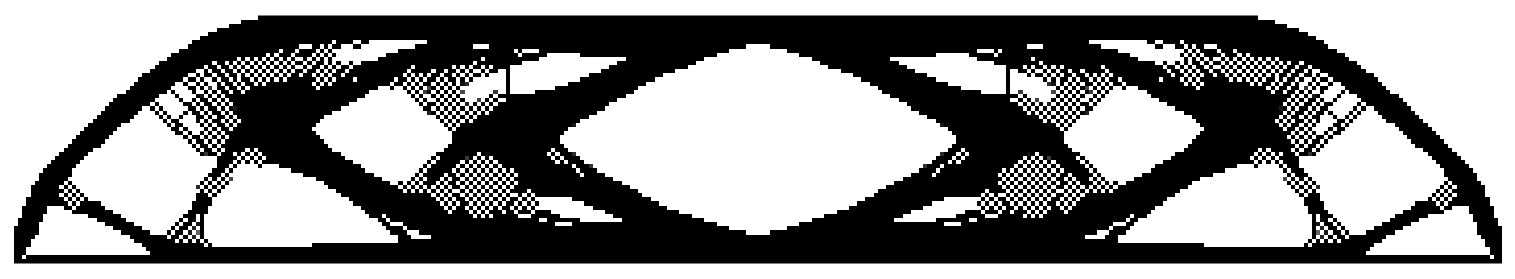}
		\caption{}
		\label{fig:quad180by60}
	\end{subfigure}
	\begin{subfigure}{0.45\textwidth}
		\centering
		\includegraphics[scale = 0.32]{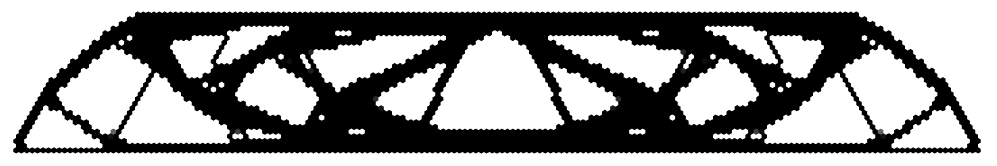}
		\caption{}
		\label{fig:hex90by30}
	\end{subfigure}
	\begin{subfigure}{0.45\textwidth}
		\centering
		\includegraphics[scale = 0.32]{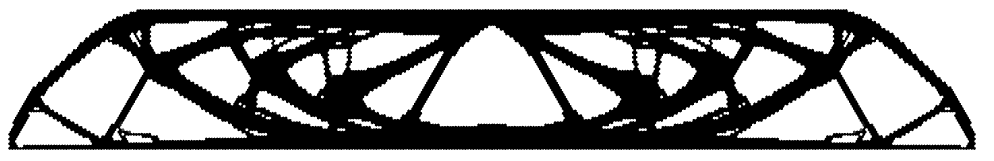}
		\caption{}
		\label{fig:hex180by60}
	\end{subfigure}
	\caption*{Figure 8: The optimized MBB beam designs. \subref{fig:quad90by30}-\subref{fig:quad180by60} results with quadrilateral FEs (\texttt{top88}), and \subref{fig:hex90by30}-\subref{fig:hex180by60} results are with hexagonal elements (\texttt{HoneyTop90}). $90\times 30$ FEs and $180\times60$ FEs are used to generate results in the first and second columns, respectively.}	\label{fig:MBBresultshexquad}
\end{figure}

Figure~8 depicts the optimized designs obtained using \texttt{top88} and \texttt{HoneyTop90} codes with different mesh sizes. One can note that the optimized designs obtained via quadrilateral FEs (Fig.~8a and Fig.~8b) contain checkerboard patterns, patches of the alternate void and solid FEs. However, such fine patches are not observed in the optimized design obtained via hexagonal FEs (Fig.~8c and Fig.~8d). Thus, honeycomb tessellation circumvents checkerboard patterns automatically due to its geometrical constructions, i.e., edge connections between two neighboring FEs. 

\section{Simple extensions}\label{Sec:SimpleExtension}
Herein, various simple extensions of the presented MATLAB code are described to solve different design problems with different input loads and boundary conditions. Heaviside projection filter scheme is also implemented in Sec.~\ref{SubSec:HeavisideProjectFilter}.
\subsection{Michell structure}\label{SubSec:MichellStructure}
\begin{figure}[h!]
	\centering
	\begin{subfigure}{0.22\textwidth}
		\centering
		\includegraphics[scale = 0.5]{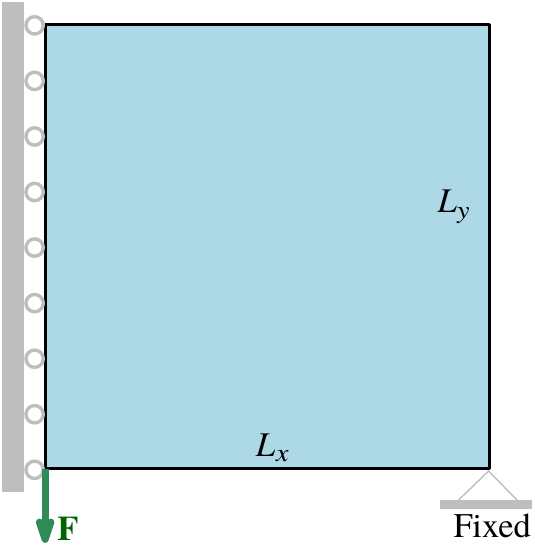}
		\caption{Michell design domain}
		\label{fig:Michelldomain}
	\end{subfigure}
	\begin{subfigure}{0.22\textwidth}
		\centering
		\includegraphics[scale = 0.27]{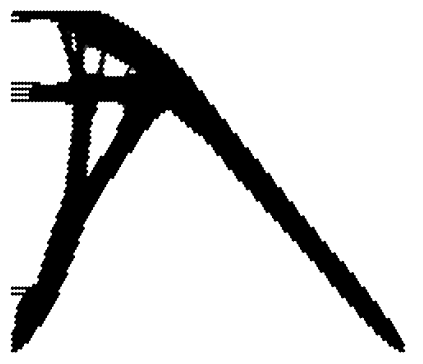}
		\caption{$\texttt{ft = 0},\, C = 58.53$}
		\label{fig:Michellft0}
	\end{subfigure}
	\begin{subfigure}{0.22\textwidth}
		\centering
		\includegraphics[scale = 0.27]{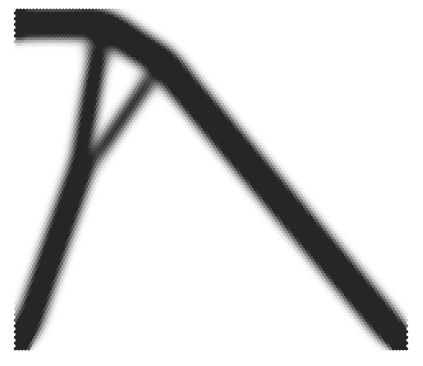}
		\caption{$\texttt{ft = 1},\, C = 59.32$}
		\label{fig:Michellft1}
	\end{subfigure}
	\begin{subfigure}{0.22\textwidth}
		\centering
		\includegraphics[scale = 0.27]{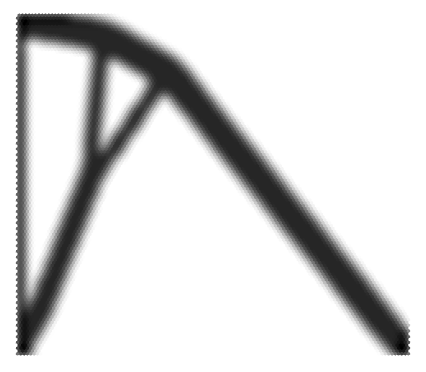}
		\caption{$\texttt{ft = 2},\, C = 83.78$}
		\label{fig:Michellft2}
	\end{subfigure}
	\caption{A symmetric half design domain for the Michell structure is shown in (\subref{fig:Michelldomain}). The corresponding load and boundary conditions are also depicted.  Optimized results (\subref{fig:Michellft0}) without filtering (\subref{fig:Michellft1}) with sensitivity filter and (\subref{fig:Michellft2}) with density filter are displayed.} \label{fig:Michellresults}
\end{figure}

We design a Michell structure  to demonstrate the code with different boundary conditions. In view of symmetry, we have only used the right half of the design domain that is depicted in Fig.~\ref{fig:Michelldomain}. The corresponding loads and boundary conditions are also shown. To accommodate this problem in the presented code, the following changes are performed:
Line~20 is altered to 
\begin{lstlisting}[numbers = none]
	F = sparse(2*1,1,-1,2*Nnode,1); % Input force
\end{lstlisting}
and line~22 is changed to
\begin{lstlisting}[numbers = none]
	fixeddofs = [2*(1:2*HNex+1:(2*HNex+1)*HNey+1)-1,(2*(2*HNex+1)), (2*(2*HNex+1))-1]; % Fixed DOFs
\end{lstlisting}
With these above modifications, we call the code as  
\begin{lstlisting}[numbers = none]
	HoneyTop90(120,120,0.20,3,3.6*sqrt(3),ft);
\end{lstlisting}
and the obtained final designs are depicted in Fig.~\ref{fig:Michellresults} for different \texttt{ft} values. One notices that  optimization with \texttt{ft=0} (Fig.~\ref{fig:Michellft0}) gives a checkerboard free optimized designs, however thin members can be seen as also noted earlier. The obtained optimized designs with $\texttt{ft=2}$ (Fig.~\ref{fig:Michellft1}) and $\texttt{ft=3}$ (Fig.~\ref{fig:Michellft2}) have different topologies. This is because, sensitivity filter ($\texttt{ft=2}$) and density filter ($\texttt{ft=3}$) have different definition (see \eqref{Eq:densityfilter} and \eqref{Eq:sensfilter}). Optimized design obtained with ft=0 is the best performing structure, and that with ft=3 is worst the performing continuum.
\begin{figure}[h!]
	\centering
	\begin{subfigure}{0.450\textwidth}
		\centering
		\includegraphics[scale = 1]{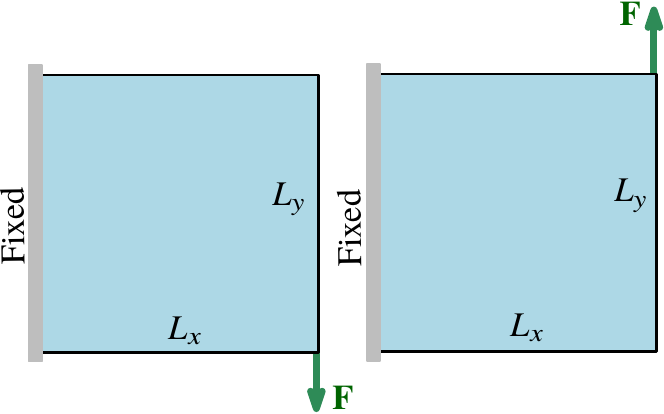}
		\caption{}
		\label{fig:Multiloaddomain}
	\end{subfigure}
	\begin{subfigure}{0.450\textwidth}
		\centering
		\includegraphics[scale = 0.30]{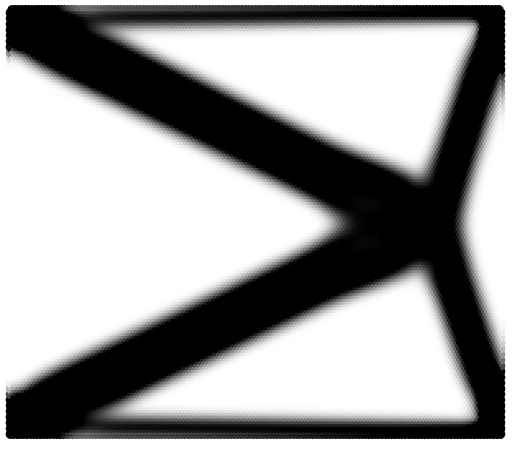}
		\caption{$\texttt{ft = 1},\,C = 86.4162$}
		\label{fig:Multiloadsolut}
	\end{subfigure}
	\caption{Design domain for a cantilever beam with two load cases is displayed in (\subref{fig:Multiloaddomain}). The optimized design is  shown in (\subref{fig:Multiloadsolut}).}\label{fig:Multiloadcases}
\end{figure}
\subsection{Multiple loads}\label{SubSec:MultipleLoads}
A cantilever beam design displayed in Fig.~\ref{fig:Multiloaddomain} \citep{sigmund200199} is solved with two load cases herein by modifying \texttt{HoneyTop90} code.

As the problem involves two load cases, the input force is placed in a two-column vector. The corresponding displacement vector is determined and  recorded in a two-column vector.  The objective function is determined as
\begin{equation}
	C = \displaystyle\sum_{k=1}^{2}\mathbf{U}_k^\top \mathbf{K} \mathbf{U}_k
\end{equation}
where $\mathbf{U}_k$ indicates displacement 
for the $k^\text{th}$ load case. In the code, lines 20, 21 and 22 are modified to
\begin{lstlisting}[numbers = none]
	F = sparse([(2*HNex+1)*2, 2*Nnode],[1 2],[-1 1],2*Nnode,2); % Input force
\end{lstlisting}
\begin{lstlisting}[numbers = none]
	U = zeros(2*Nnode,2); % Initializing displacement vector
\end{lstlisting}
and
\begin{lstlisting}[numbers = none]
	fixeddofs = [2*(1:2*HNex+1:(2*HNex+1)*HNey+1)-1,2*(1:2*HNex+1:(2*HNex+1)*HNey+1)]; % Fixed DOFs
\end{lstlisting}
respectively. $\mathbf{U}$ is determined as
\begin{lstlisting}[numbers = none]
	U(freedofs,:) = decomposition(K(freedofs,freedofs),'chol','lower')\F(freedofs,:);
\end{lstlisting}

To evaluate the objective and corresponding sensitivities, lines~64-66 are substituted by
\begin{lstlisting}[numbers = none]
	c = 0; 
	dc = 0;
	for i = 1:size(F,2)
	Ui = U(:,i); % displacement for load case i
	ce = sum((Ui(HoneyDOFss)*KE).*Ui(HoneyDOFss),2);
	c = c + sum(sum((Emin+xPhys.^penal*(E0-Emin)).*ce));
	dc = dc -penal*(E0-Emin)*xPhys.^(penal-1).*ce;
	end
\end{lstlisting}

and the 90 lines code is called as
\begin{lstlisting}[numbers = none]
	HoneyTop90(120,120,0.4,3,4*sqrt(3),1);
\end{lstlisting}
that gives the optimized design displayed in Fig.~\ref{fig:Multiloadsolut}.

\subsection{Passive design domains}\label{SubSec:PassiveDesignDomains}
\begin{figure}[h!]
	\centering
	\begin{subfigure}{0.30\textwidth}
		\centering
		\includegraphics[scale = 0.850]{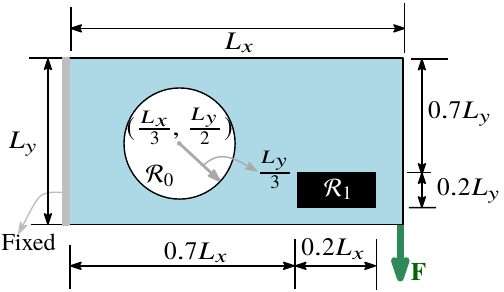}
		\caption{Design domain}\label{fig:PassActDomain}
	\end{subfigure}
	\begin{subfigure}{0.30\textwidth}
		\centering
		\includegraphics[scale = 0.380]{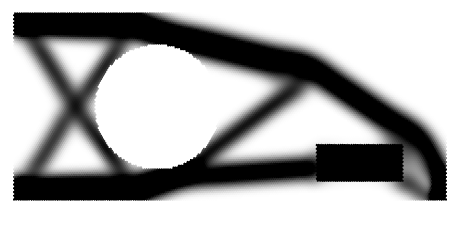}
		\caption{$\texttt{ft=1},\,C = 163.29$}\label{fig:PassActft1}
	\end{subfigure}
	\begin{subfigure}{0.30\textwidth}
		\centering
		\includegraphics[scale = 0.380]{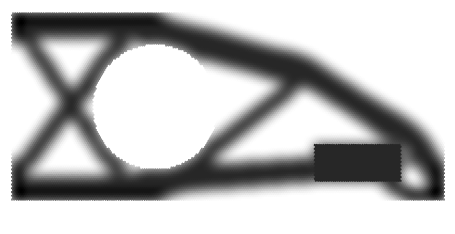}
		\caption{$\texttt{ft=2},\,C = 207.29$}\label{fig:PassActft2}
	\end{subfigure}
	\caption{A design domain with non-design solid and void regions is displayed in (\subref{fig:PassActDomain}). The optimized design with \texttt{ft=1} and \texttt{ft=2} are shown in (\subref{fig:PassActft1}) and (\subref{fig:PassActft2}) respectively.}\label{fig:PassiveDesign}
\end{figure}

In many design problems, passive (non-design)  regions characterized via  void and/or solid areas within a given design domain can exist.  For such cases, the presented code can readily be extended. The material densities of FEs associated to solid region $\mathcal{R}_1$ and void region $\mathcal{R}_0$ are fixed to 1 and 0 respectively. For example, consider the design domain shown in Fig.~\ref{fig:PassiveDesign}. The domain contains a rectangular solid region $\mathcal{R}_1$ with dimension $\frac{2L_x}{10}\times\frac{2L_y}{10}$ and a circular void area $\mathcal{R}_0$ having center at $(\frac{L_x}{3},\frac{L_y}{2})$ and radius of $\frac{L_y}{3}$. 

The presented 90-line MATLAB code is modified as follows to accommodate the problem displayed in Fig.~\ref{fig:PassActDomain}. The load vector and boundary conditions lines are changed to
\begin{lstlisting}[numbers = none]
	F = sparse(2*(2*HNex+1),1,-1,2*Nnode,1); % Input force,
\end{lstlisting}
and
\begin{lstlisting}[numbers= none]
	fixeddofs = [2*(1:2*HNex+1:(2*HNex+1)*HNey+1)-1,2*(1:2*HNex+1:(2*HNex+1)*HNey+1)]; % Fixed DOFs
\end{lstlisting}
respectively. An array $\mathtt{RSoVo}$ that contains information about FEs associated to the regions $\mathcal{R}_1$ and $\mathcal{R}_0$ (Fig.~\ref{fig:PassiveDesign}) is first initialized to $\mathtt{zeros(Nelem,1)}$ and then modified to -1 and 1 as per void and solid FEs respectively. This is performed  between lines~48-49 as 

\begin{lstlisting}[numbers = none]
	if(sqrt((ct(j,1)-max(ct(:,1))/3).^2+((ct(j,2)-max(ct(:,2))/2).^2))<max(ct(:,2))/3)
	RSoVo(j) = -1; % Non-design void region
	end
	if(ct(j,1)>0.7*max(ct(:,1))&&ct(j,1)<0.9*max(ct(:,1))&&ct(j,2)>0.1*max(ct(:,2))&&ct(j,2)<0.3*max(ct(:,2)))
	RSoVo(j) = 1; % Non-design solid region
	end
\end{lstlisting}

FEs with either -1 or 1 value in \texttt{RSoVo} remains passive during optimization, i.e., their material definition do not change with optimization evolution. The remnant elements with 0 indices in $\mathtt{RSoVo}$ act as active elements which decide the optimized material layout. To determine the active elements that are stored in the vector \texttt{act}, we introduce the following code after line~52

\begin{lstlisting}[numbers=none]
	[NVe, NSe]= deal(find(RSoVo==-1),find(RSoVo==1));
	act = setdiff((1:Nelem)', union(NVe, NSe));
\end{lstlisting}
where \texttt{NVe} and \texttt{NSe} indicate arrays of FEs associated with regions $\mathcal{R}_0$ and $\mathcal{R}_1$ respectively. The design variables associated to \texttt{act} are provided initial values as
\begin{lstlisting}[numbers = none]
	x(act) = (volfrac*(Nelem-length(NSe))-length(NVe))/length(act);
\end{lstlisting}
The sensitivities of the objective and constraints with respect to the passive design variables are zero. Therefore, while performing optimization, the active sensitivities are taken on line~77. To account non-design domains, after density filtering, the following  code on line~84 is added
\begin{lstlisting}[numbers = none]
	xPhys(RSoVo==1) = 1; xPhys(RSoVo==-1) = 0;
\end{lstlisting}

The 90-line code is called with above modifications as
\begin{lstlisting}[numbers = none]
	HoneyTop90(200,100,0.4,3,6.4*sqrt(3),ft);
\end{lstlisting}
which gives the optimized designs displayed in Fig.~\ref{fig:PassActft1} and Fig.~\ref{fig:PassActft2} using \texttt{ft=1} and \texttt{ft=2} respectively. One can note that the optimized designs have same topologies. The design obtained with ft=2 is better than that obtained with ft=3 as the former has less objective value $C$ than the that of the latter.

\subsection{Heaviside projection filter}\label{SubSec:HeavisideProjectFilter}
The Heaviside projection filter is typically employed  to achieve optimized solutions close to black-and-white~(0-1) \citep{wang2011projection}. When  a Heaviside projection filter is employed, element $j$ can be characterized via three fields--(i) design variable $\rho_j$, (ii) filtered variable $\tilde{\rho_j}$ (\ref{Eq:densityfilter}), and (iii) Heaviside projected variable $\bar{\rho_j}$ (\ref{Eq:projectionfilter}). The latter is termed physical variable herein and that is defined in terms of a filtered variable $\tilde{\rho_j}$ as \citep{wang2011projection}
\begin{equation}\label{Eq:projectionfilter}
	\bar{\rho_j}  = \frac{\tanh{\left(\beta \eta\right)} + \tanh{\left(\beta (\tilde{\rho_j}-\eta)\right)}}{\tanh{\left(\beta \eta\right)} + \tanh{\left(\beta (1-\eta)\right)}},
\end{equation}
where $\eta\in[0,\,1]$ indicates the threshold of the filter, whereas  $\beta\in[0,\,\infty)$ controls its steepness.  Typically, $\beta$ is  increased from $\beta_\text{in} =1 $ to a specified maximum value $\beta_\text{u}$ in a continuation manner. Herein, $\beta_\text{u}$ is set to 128 and $\beta$ is doubled at each 60 iterations of the optimization. The derivative of $\bar{\rho_j}$ with respect to $\tilde{\rho_j}$ is 
\begin{equation}\label{Eq:derivativeprojectionfilter}
	\pd{\bar{\rho_j}}{\tilde{\rho_j}} = \beta\frac{1-\tanh(\beta(\tilde{\rho_j} -\eta))^2}{\tanh{\left(\beta \eta\right)} + \tanh{\left(\beta (1-\eta)\right)}},
\end{equation}
and one finds derivative of $\bar{\rho_j}$ with respect to ${\rho_j}$ using the chain rule \citep{wang2011projection} and thus, derivatives of the objective and constraints.  

To accommodate this filter, the code is modified as follows. \texttt{ft=3} is used to indicate the Heaviside projection filtering steps and \texttt{move} is set to 0.1. Note that in certain cases one may have to use continuation on the move limit to control the fluctuation of the optimization process. Before the optimization loop, the following code is added
\begin{lstlisting}[numbers = none]
	beta = 1;
	eta = 0.5;
	if (ft==0|| ft == 1 || ft == 2)
	xPhys = x;
	elseif (ft == 3)
	xTilde = x; % xTilde represents filtered variables
	xPhys =  (tanh(beta*eta)+tanh(beta*(xTilde-eta)))./(tanh(beta*eta)+tanh(beta*(1-eta)));
	end
\end{lstlisting} 
To evaluate the sensitivities of objective and volume constraints these lines are introduced between lines~72-73 
\begin{lstlisting}[numbers = none]
	elseif ft==3
	dH= beta*(1-tanh(beta*(xTilde-eta)).^2)./(tanh(beta*eta)+tanh(beta*(1-eta)));
	dc = HHs'*(dc.*dH);
	dv = HHs'*(dv.*dH);
\end{lstlisting}
where vector \texttt{dH}  contains $\pd{\bar{\rho_j}}{\tilde{\rho_j}}$ \eqref{Eq:derivativeprojectionfilter}. Inside the optimization loop, between lines~81-82, we write the following
\begin{lstlisting}[numbers = none]
	elseif ft == 3
	xTilde = HHs'*x;
	xPhys = (tanh(beta*eta)+tanh(beta*(xTilde-eta)))./(tanh(beta*eta)+tanh(beta*(1-eta)));
\end{lstlisting}
and the resource constraint is employed using \texttt{xPhys}. $\beta$ is updated  in the end, between lines~89-90 as 
\begin{lstlisting}[numbers = none]
	if(ft == 3 && mod(loop,60)==0 && beta<betamax)
	beta = 2*beta;
	end
\end{lstlisting}

\begin{figure}[h!]
	\centering
	\begin{subfigure}{0.30\textwidth}
		\centering
		\includegraphics[scale = 0.38]{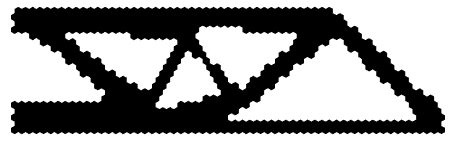}
		\caption{$C = 277.72$}
		\label{fig:Heavi6020}
	\end{subfigure}
	\begin{subfigure}{0.30\textwidth}
		\centering
		\includegraphics[scale = 0.380]{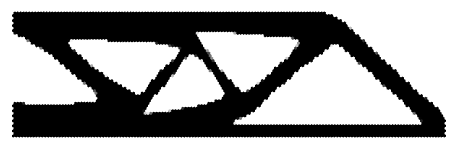}
		\caption{$C=272.29$}
		\label{fig:Heavi15050}
	\end{subfigure}
	\begin{subfigure}{0.30\textwidth}
		\centering
		\includegraphics[scale = 0.380]{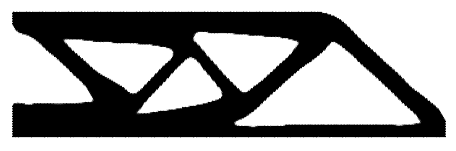}
		\caption{$C=263.10$}
		\label{fig:Heavi300100}
	\end{subfigure}
	\caption{Optimized results using \texttt{HoneyTop90} with Heaviside projection filter for the MBB beam design (Fig.~\ref{fig:MBBbeam}). Hexagonal elements used for (\subref{fig:Heavi6020}),  (\subref{fig:Heavi15050}), and  (\subref{fig:Heavi300100}) are $60\times20$, $150\times 50$ and $300 \times 100$, respectively.} \label{fig:HeavisideResults}
\end{figure}

\texttt{HoneyTop90} is called with the Heaviside projection filter for the MBB beam  design (Fig.~\ref{fig:MBBbeam}). We use the same parameters that are employed in Sec.~\ref{SubSec:MBBOptimizedResult}. For high $\beta$, sensitivity \eqref{Eq:derivativeprojectionfilter} tends to zero i.e. the filtered \texttt{dv} tends to zero and thus, \texttt{OcC} (line 77) becomes unbounded. Therefore,  the limits on Lagrange multiplier is set to $[0,\,10^9]$ i.e. a constant range, to avoid numerical instabilities as $\beta$ increases. The results are displayed in Fig.~\ref{fig:HeavisideResults} with corresponding compliance values. The obtained optimized designs contain significantly negligible number of gray elements.

\subsection{Efficiency}
	In this section, the computational cost involved in evaluating each major section of \texttt{HoneyTop90} is presented, and its overall runtime is compared with that of the 88-line code, \texttt{top88} \citep{andreassen2011efficient}.  The MBB beam design (Fig.~1) is solved for these studies. The filter radius and the volume fraction are set to $0.03$ times the length of the beam domain and 0.50 respectively. The SIMP penalty parameter $p$ (1) is set to 3.  Codes (\texttt{HoneyTop90} and \texttt{top88}) are run for 100 optimization iterations with $\texttt{ft=2}$ in MATLAB 2021a on a 64-bit operating system machine with 8.0 GB RAM, Intel(R), Core(TM) i5-8265U CPU 1.60 GHz.

The breakdown of the runtime  of \texttt{HoneyTop90} is depicted in Table~\ref{Tab:breakdowntime} for different mesh sizes. Honeycomb tessellation meshgrid information and matrices for filtering are evaluated only once for one call of \texttt{HoneyTop90}. \texttt{for} loop is used to determine filter matrices in \texttt{HoneyTop90} (line 45), therefore filter preparation time increases as mesh size grows. One can note that the meshgrid generation requires relatively negligible time (Table~\ref{Tab:breakdowntime}). FEA and optimization time increase as the mesh size increases, which is obvious.

Table~\ref{Tab:CompComp1} displays the total runtime of $\texttt{HoneyTop90}$ and \texttt{top88} for 100 optimization iterations.  We can note that $\texttt{HoneyTop90}$ performs faster than \texttt{top88} for higher mesh sizes. In \texttt{top88}, volume constraint is applied using the physical design variables (filtered designs) which are evaluated at every bisection iteration inside the optimization loop by multiplying the actual design variables to the filtering matrices whose size increase as mesh size grows. However, $\texttt{HoneyTop90}$ exploits the volume preserving nature of the density filter \citep{wang2011projection} while imposing the volume constraint and thus, reduces the overall runtime. Note also that  $\texttt{HoneyTop90}$  solves approximately double degrees of freedom (DOFs) system  to determine the displacement vector than what \texttt{top88} solves for the same mesh size (Table~\ref{Tab:CompComp1}), and takes  overall less runtime for larger mesh sizes.

\begin{table}[h!]
	\tiny
	\centering
	\caption{Breakdown of the computation time of \texttt{HoneyTop90} for 100 optimization iterations}\label{Tab:breakdowntime}
	\begin{tabular}{>{\color{black}}c >{\color{black}}c >{\color{black}}c >{\color{black}}c >{\color{black}}c >{\color{black}}c >{\color{black}}c}
		\hline \rule{0pt}{5ex}
		Mesh size                                                      & $60\times 20$ & $180\times 60$ & $300\times 100$ & $420\times 140$ & $480\times 160$ & $600\times 200$ \\ \hline \rule{0pt}{5ex}
		Meshgrid generation                                  & 0.0047 (0.089\%) & 0.0063 (0.024\%) & 0.013 (0.017\%) & 0.026 (0.013\%)  & 0.041 (0.015\%) &  0.052 (0.009\%)\\ \hline \rule{0pt}{5ex}
		Filter preparation                                   &  0.02 (0.38\%)   & 1.189 (4.65\%)  &  6.38 (8.2\%) & 26.31  (13.17\%) & 37.04 (13.06\%)  & 122.16 (21.55\%) \\ \hline \rule{0pt}{5ex}
		FEA + OC                                             & 1.73 (32.58\%)   & 19.08 (74.70\%)  &  60.01 (77.13\%) & 154.31 (77.29\%) & 215.90 (76.12\%)   &  390.736  (68.92\%)  \\ \hline \rule{0pt}{5ex}
		Plotting the solutions                               & 3.57 (67.23\%)   & 5.28 (20.67\%)  &  10.41 (13.38\%)  & 19.01 (9.53\%)  & 30.67 (10.81\%)   &  54.1  (9.54\%) \\ \hline \rule{0pt}{5ex}
		Total time of  \texttt{HoneyTop90}                   & 5.31             & 25.54           &  77.80            & 199.63          &  283.61            &  566.96  \\ \hline \rule{0pt}{5ex}
	\end{tabular}
\end{table}

\begin{table}[h!]
	\centering
	\caption{Computation time (in seconds) required by \texttt{HoneyTop90} and \texttt{top88} for 100 optimization iterations. Texts in bold indicate the lower computation time for the used mesh size.}\label{Tab:CompComp1}
	\begin{tabular}{>{\color{black}}c >{\color{black}}c >{\color{black}}c >{\color{black}}c >{\color{black}}c >{\color{black}}c >{\color{black}}c}
		\hline \rule{0pt}{5ex}
		Mesh size                                                      & $60\times 20$ & $180\times 60$ & $300\times 100$ & $420\times 140$ & $480\times 160$ & $600\times 200$ \\ \hline \rule{0pt}{5ex}
		Total DOFs for \texttt{HoneyTop90}                                    & 5080  & 44040  &  121400 & 237160  & 309440  &  482800 \\ \hline \rule{0pt}{5ex}
		Total DOFs for \texttt{top88}                                           &  2562 & 22082  &  60802 & 118722  & 154882  & 241602  \\ \hline \rule{0pt}{5ex}
		Total time of    \texttt{HoneyTop90}                     & 5.31   & 25.54   &  77.80   & \textbf{ 199.63 }   &  \textbf{283.61}   &  \textbf{566.96 }  \\ \hline \rule{0pt}{5ex}
		Total time of  \texttt{top88}                            & \textbf{3.87}  &   \textbf{ 15.92}       & \textbf{62.35 }    &  201.67  & 337.38   &  805.59  \\ \hline \rule{0pt}{5ex}
	\end{tabular}
\end{table}

\subsection{Other extensions}
The presented code can readily be extended for different set of design problems, e.g., compliant mechanisms \citep{sigmund1997design}, including heat conduction \citep{wang2011projection}, with design-dependent loads \citep{kumar2020topology,kumar2021topologySelf}, etc. One can also extend the code for the problems involving multi-physics with and without many constraints and use the  Method of Moving  Asymptotes (MMA) \citep{svanberg1987method} as an optimizer.  Extension to 3D, however, is not so straightforward, one needs to employ  tetra-kai-decahedron elements \citep{saxena2011topology} and thus, connectivity matrix and corresponding nodal coordinates are required to be generated.  

\section{Closure}\label{Sec:Closure}
This paper presents a simple, compact and efficient MATLAB code using hexagonal elements for topology optimization. The code is expected to ease the learning curve for a newcomer towards topology optimization with honeycomb tessellation.
Due to nonsingular connectivity between neighboring elements, checkerboard patterns and point connections are circumvented inherently. However, thin members are present in the optimized designs which are noticed mesh dependent. Therefore, to circumvent mesh dependence nature of hexagonal FEs in TO, one requires to use filtering techniques or other suppressing approaches.

A novel honeycomb tessellation generation approach is presented. The code generates meshgrid information, i.e., the element connectivity matrix and nodal coordinates array for the millions of hexagonal elements within a fraction of a second using the MATLAB inbuilt functions. The element connectivity matrix and corresponding nodal coordinates generation require just 5(7) and 4(6) lines. Wachpress shape functions are employed to evaluate the stiffness matrix of a hexagonal element. The optimality criteria approach is employed for  the compliance minimization problems. The provided MATLAB code (Appendix~\ref{app:MATLABcode}) and its extensions are explained in detail. Easy and efficient meshgrid generation for tetra-kai-decahedron elements, performing finite element analysis and optimization form a future direction for a three-dimensional problem setting. In addition, extensions of code to solve advanced design problems with stress and buckling constraints may be one of the prime directions for future work. 

	\section*{Acknowledgment}
	The author would like to thank Prof. Anupam Saxena, Indian Institute of Technology Kanpur, India, for fruitful discussions and  acknowledge financial support from the Science \& Engineering research board, Department of Science and Technology, Government of India under the project file number RJF/2020/000023.
	\section*{Conflicts of interest}
	None.
	\nopagebreak
	\begin{appendices}
		\onecolumn
		\numberwithin{equation}{section}
		\numberwithin{figure}{section}
		\section{HoneyTop90 MATLAB code}\label{app:MATLABcode}
\begin{lstlisting}
function HoneyTop90(HNex,HNey,volfrac,penal,rfill,ft)
%% ---------- Material properties ----------
E0 = 1;
Emin = E0*1e-9;
%% ---Element connectivity, nodal coordinates, Finite element analysis preparation--
NstartVs = reshape(1:(1+2*HNex)*(1+HNey),1+2*HNex,1+HNey);
DOFstartVs = reshape(2*NstartVs(1:end-1,1:end-1)-1,2*HNex*HNey,1);
NodeDOFs = repmat(DOFstartVs,1,8) + repmat([2*(2*HNex+1) + [2 3 0 1] 0 1 2 3 ],2*HNex*HNey,1);
ActualDOFs = NodeDOFs(setdiff(1:2*HNex*HNey,(2*HNex:2*HNex:2*HNex*HNey)' +  mod(1:HNey,2)'),:);
HoneyDOFs = [ActualDOFs(2:2:end,1:2), ActualDOFs(1:2:end,:), ActualDOFs(2:2:end,7:8)];
Ncyi = repmat(reshape(repmat([-0.25 0.25]',HNey+1,1),2,HNey+1)+reshape(1.5*sort(repmat((0:HNey)',2,1)),2,(HNey+1)),HNex+1,1);
Ncyi(:,1:2:end) = flip(Ncyi(:,1:2:end));
Ncyf = Ncyi(1:end-1,:); % final arrays containing y-coordinates
HoneyNCO=(1)*[repmat((0:cos(pi/6):2*HNex*cos(pi/6)),1,HNey+1)' Ncyf(:)];%node co
if(mod(HNey,2)==0)
 HoneyDOFs(end:-1:end-(HNex)+2,1:6) = HoneyDOFs(end:-1:end-(HNex)+2,1:6)-2; % Updating
 HoneyNCO([(2*HNex+1)*HNey+1;(2*HNex+1)*(HNey+1)],:) = []; % Removing hangining nodes
end
[Nelem,Nnode] = deal(size(HoneyDOFs,1),size(HoneyNCO,1)); % elem #, node #
 F = sparse(2*((2*HNex+1)*HNey+1),1,-1,2*Nnode,1);        % Applied load
 U = zeros(2*Nnode,1);                                    % Initializing U
 fixeddofs = [2*(1:2*HNex+1:(2*HNex+1)*HNey+1)-1,(2*(2*HNex+1))];  % Fixed DOFs
 alldofs = 1:2*Nnode;                                              % Total DOFs
 freedofs = setdiff(alldofs,fixeddofs);                            % Free DOFs
 iK = reshape(kron(HoneyDOFs,ones(12,1))',144*Nelem,1);
 jK = reshape(kron(HoneyDOFs,ones(1,12))',144*Nelem,1);
 KE = E0*[616.43012 92.77147 -168.07333 65.54377 -232.28511 -0.00032 -120.65312 -83.28564 -71.60020 -92.77115 -23.81836 17.74187;
 92.77147 509.30685 101.02751 -71.90335 0.00032 -18.03857 -83.28564 -24.48314 -92.77179 -178.72347 -17.74187 -216.15832;
-168.07333 101.02751 455.74522 0.00000 -168.07333 -101.02751 -71.60020 -92.77179 23.60185 -0.00000 -71.60020 92.77179;
65.54377 -71.90335 0.00000 669.99176 -65.54377 -71.90335 -92.77115 -178.72347 -0.00000 -168.73811 92.77115 -178.72347;
-232.28511 0.00032 -168.07333 -65.54377 616.43012 -92.77147 -23.81836 -17.74187 -71.60020 92.77115 -120.65312 83.28564;
-0.00032 -18.03857 -101.02751 -71.90335 -92.77147 509.30685 17.74187 -216.15832 92.77179 -178.72347 83.28564 -24.48314;
-120.65312 -83.28564 -71.60020 -92.77115 -23.81836 17.74187 616.43012 92.77147 -168.07333 65.54377 -232.28511 -0.00032;
-83.28564 -24.48314 -92.77179 -178.72347 -17.74187 -216.15832 92.77147 509.30685 101.02751 -71.90335 0.00032 -18.03857;
-71.60020 -92.77179 23.60185 -0.00000 -71.60020 92.77179 -168.07333 101.02751 455.74522 0.00000 -168.07333 -101.02751;
-92.77115 -178.72347 -0.00000 -168.73811 92.77115 -178.72347 65.54377 -71.90335 0.00000 669.99176 -65.54377 -71.90335;
-23.81836 -17.74187 -71.60020 92.77115 -120.65312 83.28564 -232.28511 0.00032 -168.07333 -65.54377 616.43012 -92.77147;
17.74187 -216.15832 92.77179 -178.72347 83.28564 -24.48314 -0.00032 -18.03857 -101.02751 -71.90335 -92.77147 509.30685;]/1000;% elem stiffness 
%% ---------- Filter preperation ---------
Cxx = repmat([sqrt(3)/2*(1:2:2*HNex-1) sqrt(3)*(1:1:HNex-1)],1,ceil(HNey/2));
Cyy= (repmat(3/4,HNex,HNey) + repmat(3/2*(0:HNey-1),HNex,1));
Cyy(HNex+1:2*HNex:length(Cyy(:))) = [];
ct = [Cxx(1:length(Cyy))' Cyy']*(1);             % Centre coordinates
DD = cell(Nelem,1);                                      % Initializing 
for j = 1:Nelem
    Cent_dist = sqrt((ct(j,1)-ct(:,1)).^2+((ct(j,2)-ct(:,2)).^2));
    [I,J] = find(Cent_dist<=rfill);
    DD{j} = [I,J+(j-1),Cent_dist(I)];
end
 DD = cell2mat(DD);
 HHs = sparse(DD(:,2),DD(:,1),1-DD(:,3)/rfill);
 HHs = spdiags(1./sum(HHs,2),0,Nelem,Nelem)*HHs;                           
%% ---------- Initialization -------------
x = volfrac*ones(Nelem,1);                                               % Initial guess
[xPhys,loop,change,maxiter,dv,move] = deal(x,0,1,200,ones(Nelem,1),0.2); % Parameters 
%% ---------- Start optimization ----------
while (change > 0.01 && loop < maxiter)
  loop = loop + 1;
  %% ---------- Finite element analysis ----------
  sK = reshape(KE(:)*(Emin + xPhys'.^penal*(E0-Emin)),144*Nelem,1);
  K = sparse(iK,jK,sK);                                            % Global stiffness 
  U(freedofs) = decomposition(K(freedofs,freedofs),'chol','lower')\F(freedofs);
  %% ---------- Objective and sensitivities evaluation ----------
  ce = sum((U(HoneyDOFs)*KE).*U(HoneyDOFs),2);
  c =  sum(sum((Emin+xPhys.^penal*(E0-Emin)).*ce));               % Finding objective 
  dc = -penal*(E0-Emin)*xPhys.^(penal-1).*ce;                     % Obj. sensitivities
   %% ---------- Using Filters ---------------------------
   if ft == 1
    dc = HHs'*(x.*dc)./max(1e-3,x);
  elseif ft == 2
    dc = HHs'*(dc);
    dv = HHs'*(dv);
   end
  %% ---------- Optimality criteria update ----------------
   xOpt = x;
  [xUpp, xLow] = deal (xOpt + move, xOpt - move);              % Upp. & low. limits
  OcC = xOpt.*(sqrt(-dc./dv));                                 % Opt. parameter 
  inL = [0, mean(OcC)/volfrac];                                % Lag. Mul. range
  while (inL(2)-inL(1))/(inL(2)+inL(1))> 1e-3
    lmid = 0.5*(inL(2)+ inL(1));
    x = max(0,max(xLow,min(1,min(xUpp,OcC/lmid))));
    if mean(x)>volfrac, inL(1) = lmid; else, inL(2) = lmid; end 
  end
  if ft == 1 || ft ==0, xPhys = x; elseif ft == 2, xPhys = HHs'*x; end
  change = max(abs(xOpt-x));
  %% ---------- Results printing ----------------------------
  fprintf(' It.:%5i Obj.:%11.4f Vol.:%7.3f ch.:%7.3f\n',loop,c, mean(xPhys),change);
  %% ---------- Plotting intermediate designs ---------------
  colormap('gray'); scatter(ct(:,1),ct(:,2),[],1-xPhys,'filled'); axis off equal; pause(1e-6);
end
\end{lstlisting}
		\section{Sensitivity analysis} \label{append:sensanalysis}
In this paper, the optimality criteria approach is employed for the optimization. Therefore, derivatives of the objective and constraint with respect to the design variables are required. Herein, the adjoint-variable method is used to determine the sensitivity of the objective, $C({\bm{\rho}}) = \mathbf{u}^\top \mathbf{K}(\bm{\rho})\mathbf{u}$. The overall performance function $\mathcal{L}$ in conjunction with the state equation, $\mathbf{K} \mathbf{u} - \mathbf{F} = \mathbf{0}$ is written as
\begin{equation}\label{eq:performancefunction}
	\mathcal{L} = C + \bm{\lambda}^\top (\mathbf{K} \mathbf{u} - \mathbf{F}),
\end{equation}
where $\bm{\lambda}$  is the Lagrange multiplier vector.
In view of \eqref{eq:performancefunction}, one finds derivative of $\mathcal{L}$ with respect to $\bm{\rho}$ as

\begin{equation}\label{eq:performderivative}
	\begin{split}
		\frac{\partial\mathcal{L}}{\partial \bm{\rho}} &= \pd{C}{\bm{\rho}} + \pd{C}{\mathbf{u}}\pd{\mathbf{u}}{\bm{\rho}} + \bm{\lambda}^\top \left(\pd{\mathbf{K}}{\bm{\rho}} \mathbf{u}+ \mathbf{K} \pd{\mathbf{u}}{\bm{\rho}}\right)\\
		& = \mathbf{u}^\top\pd{\mathbf{K}}{\bm{\rho}}\mathbf{u} + 2\mathbf{u}^\top\mathbf{K}\pd{\mathbf{u}}{\bm{\rho}} +  \bm{\lambda}^\top\pd{\mathbf{K}}{\bm{\rho}}\mathbf{u} + \bm{\lambda}^\top\mathbf{K} \pd{\mathbf{u}}{\bm{\rho}} \qquad \left(\text{using},\, C({\bm{\rho}}) = \mathbf{u}^\top \mathbf{K}(\bm{\rho})\mathbf{u}\right)\\
		& =  \mathbf{u}^\top\pd{\mathbf{K}}{\bm{\rho}}\mathbf{u} +  \bm{\lambda}^\top\pd{\mathbf{K}}{\bm{\rho}}\mathbf{u} + \underbrace{\left( 2\mathbf{u}^\top\mathbf{K} + \bm{\lambda}^\top\mathbf{K}\right)}_{\Theta}\pd{\mathbf{u}}{\bm{\rho}}
	\end{split}
\end{equation}
In \eqref{eq:performderivative}, $\Theta = 0$, the adjoint equation, yields $\bm{\lambda} = -2\mathbf{u}$ and thus, one writes \eqref{eq:performancefunction} as
\begin{equation}\label{eq:performderivativef}
	\begin{split}
		\frac{\partial\mathcal{L}}{\partial \bm{\rho}} 
		& =  \mathbf{u}^\top\pd{\mathbf{K}}{\bm{\rho}}\mathbf{u}   -2\mathbf{u}^\top\pd{\mathbf{K}}{\bm{\rho}}\mathbf{u} \\	
		& = - \mathbf{u}^\top\pd{\mathbf{K}}{\bm{\rho}}\mathbf{u}
	\end{split}
\end{equation}
Therefore, in view of \eqref{eq:performderivativef}, the derivative of objective $C$ with respect to design variable $\rho_j$ can be written as 
\begin{equation}
	\pd{C}{\rho_j} = -\mathbf{u}_j^\top\frac{\partial \mathbf{k}_j}{\partial \rho_j} \mathbf{u}_j 
\end{equation}
where $\mathbf{u}_j$ and $\mathbf{k}_j$ are the displacement vector and the stiffness matrix of element $j$, respectively.
			\section{Wachspress shape functions}\label{appsec:Wachspress_shape_functions}
\begin{figure}[h!]
	\centering
	\includegraphics[scale = 0.75]{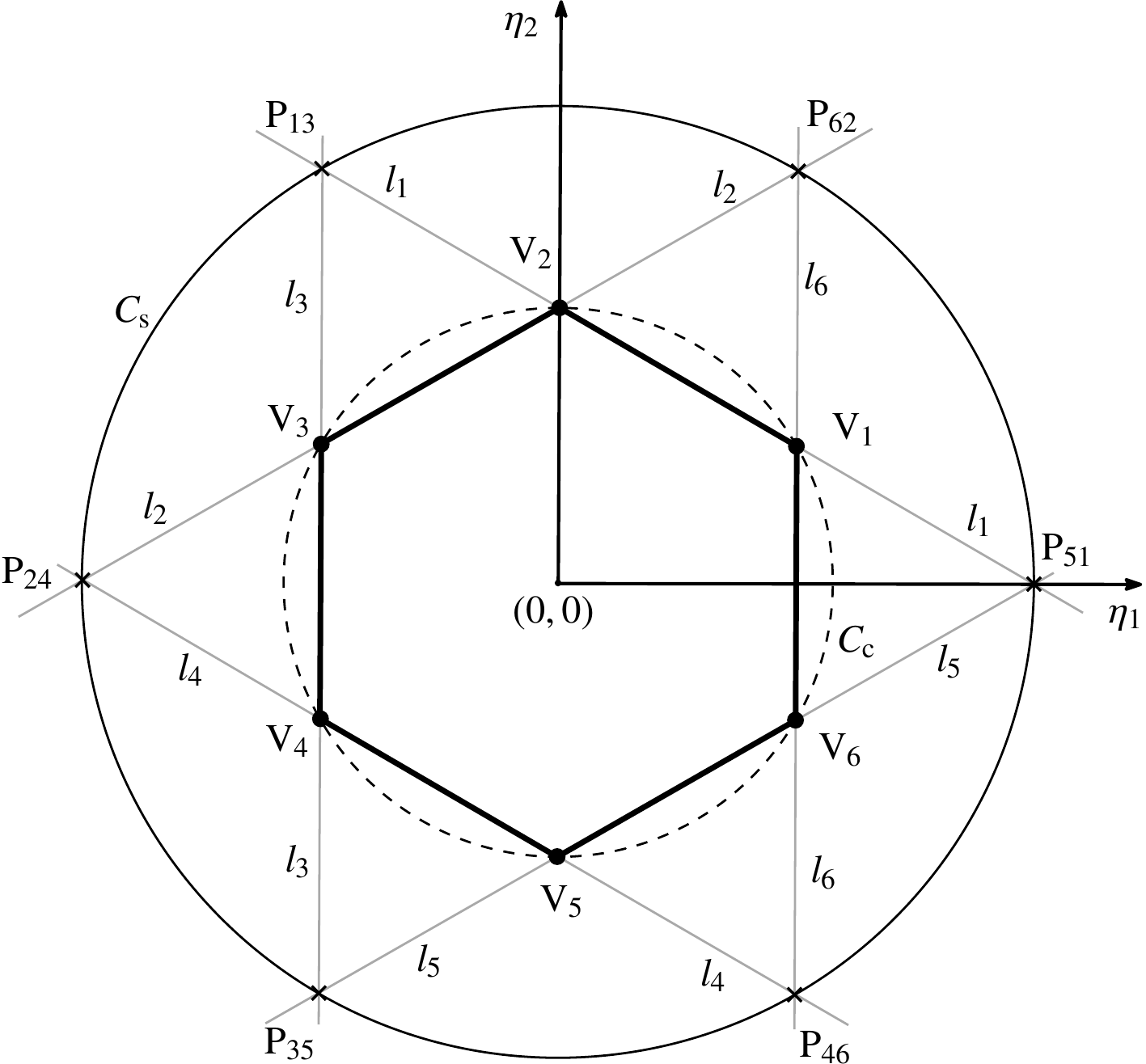}
	\caption{A regular hexagonal element with vertices V$_i|_{i = 1,\,2,\,\cdots,\,6}$ and circumscribing circle ${C}_\text{c}$ with radius 1 unit. Coordinates of vertex V$_i$ are $\left((\eta_1^i,\,\eta_2^i) \equiv \left(\cos(\frac{(2i-1)\pi}{6}),\,\sin(\frac{(2i-1)\pi}{6})\right)\right)$. Straight lines $l_i$ pass through vertices V$_i$ and V$_{i-1}$. P$_{i\,{i+2}}$ are the intersection points of straight lines $l_i$ and $l_{i+2}$. Circle ${C}_\text{s}$ with radius  $\sqrt{3}$ unit  is drawn that passes through points P$_{i\,{i+2}}$.}	\label{fig:shapefunctions}
\end{figure}
Figure~\ref{fig:shapefunctions} depicts a hexagonal element with vertices V$_i|_{i = 1,\,2,\,3,\,\cdots,\,6}$ in $\bm{\eta}$~co-ordinates system.  Coordinates of vertex V$_i$ are $\left((\eta_1^i,\,\eta_2^i) \equiv \left(\cos(\frac{(2i-1)\pi}{6}),\,\sin(\frac{(2i-1)\pi }{6})\right)\right)$.  The circumscribing circle with radius~1~unit is represented via ${C}_\text{c}$. Let  Wachspress shape function for vertex V$_i$ (Fig.~\ref{fig:shapefunctions}) be $N_i$. Using the fundamentals of coordinate geometry and in view of coordinates of V$_i$, the equations of straight lines $l_i$ (Fig.~\ref{fig:shapefunctions}) can be written as
\begin{equation}\label{eq:lineequations}
	\begin{rcases}
		l_1(\bm{\eta}) &\equiv \eta_1 + \sqrt{3}\eta_2 -\sqrt{3} = 0\\
		l_2(\bm{\eta}) &\equiv  -\eta_1 + \sqrt{3}\eta_2 -\sqrt{3} = 0 \\
		l_3(\bm{\eta}) &\equiv 2\eta_1 +\sqrt{3} = 0\\
		l_4(\bm{\eta}) &\equiv \eta_1 + \sqrt{3}\eta_2 +\sqrt{3} = 0\\
		l_5(\bm{\eta}) &\equiv -\eta_1 + \sqrt{3}\eta_2 +\sqrt{3} = 0\\
		l_6(\bm{\eta}) &\equiv 2\eta_1 -\sqrt{3} = 0 
	\end{rcases},
\end{equation}
and likewise, the equation of circle ${C}_\text{s}$~(cf. Fig.~\ref{fig:shapefunctions}, passing through P$_{i i+2}$) can be written as
\begin{equation}\label{eq:circleequation}
	{C}_\text{s}(\bm{\eta}) \equiv \eta_1^2 + \eta_2^2 - 3 = 0.
\end{equation}
Straight lines $l_i$ and $l_{i+2}$ intersect at points P$_{i\,{i+2}}$~(Fig.~\ref{fig:shapefunctions}). The shape function of node~1, i.e., $N_1$ is determined as \citep{wachspress1975rational}
\begin{equation}\label{eq:shapefunction_i}
	N_1 = s_1\frac{l_2(\bm{\eta}) l_3(\bm{\eta}) l_4(\bm{\eta}) l_5(\bm{\eta})}{{C}_\text{s}(\bm{\eta})},
\end{equation}
where $s_1$, a constant, is calculated using the Kronecker-delta property of a shape function, which is defined as

\begin{equation}\label{eq:Kronecker delta}
	N_i(\bm{\eta}_j) = \delta_{ij} =
	\begin{lcases}
		1, \quad \text{if}\,\, i = j\\
		0, \quad \text{if}\,\,i\ne j\\
	\end{lcases}.
\end{equation}
Now, in view of coordinates\footnote{$\eta_1^1=\cos(\frac{\pi}{6}),\,\eta_2^1=\sin(\frac{\pi}{6})$} of node $1$, i.e., $\left(\eta_1^1,\,\eta_2^1\right)$ and \eqref{eq:Kronecker delta}, \eqref{eq:shapefunction_i} yields
\begin{equation}
	s_1 = \frac{{C}_\text{s}(\eta_1^1,\,\eta_2^1)}{l_2(\eta_1^1,\,\eta_2^1) l_3(\eta_1^1,\,\eta_2^1) l_4(\eta_1^1,\,\eta_2^1) l_5(\eta_1^1,\,\eta_2^1)} = \frac{1}{18}.
\end{equation}
Likewise, one can determine Wachspress shape functions for all other nodes with their respective constants. The final expressions of the shape functions are:
\begin{equation}
	\begin{rcases}
		N_1(\bm{\eta}) &= \frac{\left(-\eta_1 + \sqrt{3}\eta_2 -\sqrt{3}\right)\left(2\eta_1 +\sqrt{3}\right)\left(\eta_1 + \sqrt{3}\eta_2 +\sqrt{3}\right)\left(-\eta_1 + \sqrt{3}\eta_2 +\sqrt{3}\right)}{18\left(\eta_1^2 + \eta_2^2 - 3\right)}\\
		N_2(\bm{\eta}) &=\frac{\left(2\eta_1 +\sqrt{3}\right)\left(\eta_1 + \sqrt{3}\eta_2 +\sqrt{3}\right)\left(-\eta_1 + \sqrt{3}\eta_2 +\sqrt{3}\right)\left(2\eta_1 -\sqrt{3}\right)}{18\left(\eta_1^2 + \eta_2^2 - 3\right)}\\
		N_3(\bm{\eta}) &= -\frac{\left(\eta_1 + \sqrt{3}\eta_2 +\sqrt{3}\right)\left(-\eta_1 + \sqrt{3}\eta_2 +\sqrt{3}\right)\left(2\eta_1 -\sqrt{3}\right)\left(\eta_1 + \sqrt{3}\eta_2 -\sqrt{3}\right)}{18\left(\eta_1^2 + \eta_2^2 - 3\right)}\\
		N_4(\bm{\eta}) &= \frac{\left(-\eta_1 + \sqrt{3}\eta_2 +\sqrt{3}\right)\left(2\eta_1 -\sqrt{3}\right)\left(\eta_1 + \sqrt{3}\eta_2 -\sqrt{3}\right)\left(-\eta_1 + \sqrt{3}\eta_2 -\sqrt{3}\right)}{18\left(\eta_1^2 + \eta_2^2 - 3\right)}\\
		N_5(\bm{\eta}) &= \frac{\left(2\eta_1 -\sqrt{3}\right)\left(\eta_1 + \sqrt{3}\eta_2 -\sqrt{3}\right)\left(-\eta_1 + \sqrt{3}\eta_2 -\sqrt{3}\right)\left(2\eta_1 +\sqrt{3} \right)}{18\left(\eta_1^2 + \eta_2^2 - 3\right)}\\
		N_6(\bm{\eta}) &= -\frac{\left(\eta_1 + \sqrt{3}\eta_2 -\sqrt{3}\right)\left(-\eta_1 + \sqrt{3}\eta_2 -\sqrt{3}\right)\left(2\eta_1 +\sqrt{3}\right)\left( \eta_1 + \sqrt{3}\eta_2 +\sqrt{3}\right)}{18\left(\eta_1^2 + \eta_2^2 - 3\right)}\\
	\end{rcases}.
\end{equation} 
\section {Numerical Integration}\label{appsec:Numerical_quadrature_points}
To evaluate stiffness matrix of an element, numerical integration approach using the quadrature points is employed. As per \citep{lyness1977quadrature}, quadrature points  for a hexagonal element are given in Table~\ref{tab:quadraturepoint} \citep{saxena2013combined}, and a function $f\left(\eta_1,\,\eta_2\right)$ can be integrated as
\begin{equation}
	\int_{A_6}f\left(\eta_1,\,\eta_2\right) \text{d}\Omega \approx A_6\left(w_o f(0,\,0) + \sum_{k=2}^{k_\text{max}}\sum_{i=1}^{6} w_k f\left(r_k,\,\alpha_k + \frac{\pi i}{3}\right)\right).
\end{equation}
where $A_6$ is the area of a regular hexagonal element. Table~\ref{tab:quadraturepoint} notes the quadrature points $N$ with coordinates $\left(\eta_1^a,\,\eta_2^a\right)$ $ = \left(r_k\cos(\alpha_k + \frac{i\pi}{3}),r_k\sin(\alpha_k + \frac{i\pi}{3})\right)$. $w_k$ indicate the weights for these points.  $i =1,\,2,\,3,\,\cdots,\,6$, if $k>1$, and $ i =1,\,\text{if}\, k=1$, corresponds to single integration point at center $(0,\,0)$. For $k>1$, six integration points lie on the a circle with center at (0,\,0) and radius $r_k$. Note that the quadrature rule is invariant under a rotation of $60^o$ for a hexagonal element.
\begin{table}[h!]
	\centering
	\caption{Quadrature points for a hexagonal element}\label{tab:quadraturepoint}
	\begin{tabular}{c| c c c l}
		\hline \hline\rule{0pt}{3ex}
		Cases                                                                   & $r_k$              & $\alpha_k$        & $w_k$             &  \rule{0pt}{4ex} \\ \hline \rule{0pt}{3ex}
		\multirow{2}{*}{$1$,\,\quad $k = 2$,\,\quad $N = 1+6(k-1)=7$} & $0.0000$             & $0.0000$             & $0.255952380952381$ &  \\ \rule{0pt}{4ex} 
		& $0.748331477354788$ & $0.0000$             & 0.124007936507936 &  \rule{0pt}{4ex} \\ \hline\rule{0pt}{3ex}
		\multirow{3}{*}{$2$,\,\quad $k = 3$,\,\quad $N = 1+6(k-1)= 13$}             & 0.0000             & $0.0000$             & $0.174588684325077$ &   \rule{0pt}{4ex} \\ 
		& $0.657671808727194$ & $0.0000$            & $0.115855303626943$ &  \rule{0pt}{4ex} \\ 
		& $0.943650632725263$ & $0.523681372148045$ & $0.021713248985544$ &  \rule{0pt}{4ex} \\ \hline\rule{0pt}{3ex}
		\multirow{4}{*}{$3$,\,\quad $k = 4$,\,\quad $N = 1+6(k-1)= 19$}             & 0.0000            & $0.0000$            & $0.110826547228661$ &  \rule{0pt}{4ex} \\ 
		& $0.792824967172091$ & $0.0000$            & $0.037749166510143$ &  \rule{0pt}{4ex} \\ 
		& $0.537790663359878$ & $0.523598775598299$ & $0.082419705350590$ &  \rule{0pt}{4ex} \\ 
		& $0.883544457934942$ & $0.523598775598299$ & $0.028026703601157$ &  \rule{0pt}{4ex} \\ \hline\rule{0pt}{3ex}
		\multirow{5}{*}{$4$,\,\quad $k = 5$,\,\quad $N= 1+6(k-1)= 25$}              & $0.0000$            & $0.0000$            & $0.087005549094808$ &  \rule{0pt}{4ex} \\ 
		& $0.487786213872069$ & $0.0000$            & $0.071957468118574$ &  \rule{0pt}{4ex} \\ 
		& $0.820741657108524$ & $0.0000$            & $0.027500185650866$ &  \rule{0pt}{4ex} \\ 
		& $0.771806696813652$ & $0.523598775598299$ & $0.045248932131663$ &  \rule{0pt}{4ex} \\ 
		& $0.957912268790000$ & $0.523598775598299$ & $0.007459892497607$ &  \rule{0pt}{4ex} \\ \hline \hline
	\end{tabular}
\end{table}

	\end{appendices}
	\bibliography{myreference}
	\bibliographystyle{spbasic} 
\end{document}